\documentclass[prd,nofootinbib,twocolumn,a4paper,showpacs,showkeys,
preprintnumbers]{revtex4}

\usepackage{graphicx}
\usepackage{amsmath}
\usepackage{amssymb}
\usepackage{amsfonts}
\usepackage{latexsym}
\usepackage{mathrsfs}

\newcommand{\tr}{{\rm tr}}
\newcommand{\Tr}{{\rm Tr}}
\newcommand{\sign}{{\rm sign}}
\renewcommand{\Re}{{\rm Re}}
\renewcommand{\Im}{{\rm Im}}

\newcommand{\intg}{{\textstyle\int}\,}
\newcommand{\CP}{\textit{CP}}
\newcommand{\dg}{\mathscr{D}^4}
\renewcommand{\vec}[1]{{\bf #1}}
\newcommand{\textint}{\textstyle\int}

\newcommand{\momp}{{p}_2}
\newcommand{\momq}{{p}_3}
\newcommand{\momr}{{p}_4}

\newcommand{\momabs}{|\vec{p}|}
\newcommand{\momkabs}{|\vec{p}|}
\newcommand{\mompabs}{|\vec{p}_2|}
\newcommand{\momqabs}{|\vec{p}_3|}
\newcommand{\momrabs}{|\vec{p}_4|}

\newcommand{\bvec}[1]{\vec{#1}}

\newcommand{\lorig}{L}

\newcommand{\carrayorigkb}[3]{{C}_{#1}\ifeqthenelse{}{#3}{}{[{#3}]}\ifeqthenelse
{}{#2}{}{(#2)}}

\newcommand{\qstat}[1]{[1+#1]}

\newcommand{\hubblerate}{{H}}

\newcommand{\Cmasspsi}{10^{10}}

\newcommand{\Cdim}{400}

\newcommand{\Ckappaa}{0.01}
\newcommand{\Ckappab}{0.1}
\newcommand{\Ckappac}{0.366}
\newcommand{\Ckappad}{1}
\newcommand{\Ckappae}{10}
\newcommand{\Ckappaf}{100}

\newcommand{\Ckappamaxeta}{0.059}
\newcommand{\Ckappacmaxreleta}{0.34}

\newcommand{\ifeqthenelse}[4]{\edef\tempa{#1}\def\tempb{#2}\ifx\tempa\tempb{#3}
\else{#4}\fi}

\newcommand{\dif}{d}

\providecommand{\bbar}{\bar{b}}
\newcommand{\f}[2]{f_{#1}\ifeqthenelse{}{#2}{}{(#2)}}
\providecommand{\fig}{Fig.\,}
\providecommand{\Fig}{Fig.\,}

\providecommand{\eqn}{Eq.\,}
\providecommand{\eqns}{Eqs.\,}

\providecommand{\dend}{{\, .}}

\providecommand{\kend}{{\, ,}}

\providecommand{\lorentzd}[2]{\dif\Pi^{#1}_{#2}}

\begin{document}


\preprint{TUM-HEP-740/09}

\title{Systematic approach to leptogenesis in nonequilibrium QFT:
\\ self-energy contribution to the \textit{CP}-violating parameter}

\author{M. Garny$^{b}$}
\email[\,]{mathias.garny@ph.tum.de}

\author{A. Hohenegger$^{a}$}
\email[\,]{andreas.hohenegger@mpi-hd.mpg.de}

\author{A. Kartavtsev$^{a}$}
\email[\,]{alexander.kartavtsev@mpi-hd.mpg.de}  

\author{M. Lindner$^{a}$}
\email[\,]{manfred.lindner@mpi-hd.mpg.de}

\affiliation{%
$^a$Max-Planck-Institut f\"ur Kernphysik, Saupfercheckweg 1, 69117 Heidelberg,
Germany\\
$^b$Technische Universit\"at M\"unchen, James-Franck-Stra\ss e, 85748 Garching,
Germany}

\begin{abstract}
In the baryogenesis via leptogenesis scenario the self-energy contribution
to the \CP-violating parameter plays a very important role. Here, we calculate 
it in a simple toy model of leptogenesis using the 
Schwinger--Keldysh/Kadanoff--Baym formalism as  starting point. We show that  
the formalism is free of the double-counting problem typical for the canonical 
Boltzmann approach. Within the toy model, medium effects increase the 
\CP-violating parameter. In contrast to results obtained earlier in the 
framework of thermal field theory, the medium corrections are  linear in the 
particle number densities. In the resonant regime quantum corrections  lead to
modified expressions for the \CP-violating parameter and for the  decay width.
Most notably, in the maximal resonant regime the 
Boltzmann picture breaks down and an analysis in the full Kadanoff--Baym 
formalism is required.
\end{abstract}

\maketitle

% =============================================================================
\section{\label{introduction}Introduction}
% =============================================================================

From the theoretical point of view the baryogenesis via leptogenesis 
scenario \cite{Fukugita:1986hr} is a very attractive explanation of
the observed baryon asymmetry of the Universe. One of its key ingredients 
is heavy Majorana neutrinos, whose \textit{CP}--violating decays are 
responsible for the generation of a lepton asymmetry, i.e.~for 
leptogenesis. The \textit{CP}-violating parameter receives contributions 
from the vertex \cite{Fukugita:1986hr,Garny:2009rv} and self-energy
\cite{Flanz:1994yx,Covi:1996wh,PhysRevD.56.5431,Pilaftsis:2003gt} diagrams. If
the masses of the heavy neutrinos are strongly hierarchical, then the two
contributions are comparable. However, if the mass spectrum is quasidegenerate
then the self-energy contribution is resonantly enhanced and becomes considerably
larger than the one from the vertex diagram. The resonant enhancement of the
\CP-violating parameter allows one to bring the scale of leptogenesis down to
$\sim 1$ TeV \cite{Pilaftsis:2003gt,Pilaftsis:2005rv}. This scenario is very
interesting from the experimental point of view since it is potentially 
accessible in  accelerator experiments \cite{Blanchet:2009bu}. 

In state-of-the-art calculations the self-energy \CP-violating parameter is
evaluated in \textit{vacuum} and then used to calculate the asymmetry generated
by the  decays of the heavy neutrinos in the \textit{hot and dense medium}. In
this approximation the possibly important medium effects are neglected from the
very beginning. One can take them  into account systematically by using the
Schwinger--Keldysh/Kadanoff--Baym formalism \cite{KB:1962} or  approximate
self-consistent equations derived from the Kadanoff--Baym equations (see
\cite{Covi:1997dr,Buchmuller:2000nd,Giudice:2003jh,Lindner:2005kv,
Lindner:2007am,DeSimone:2007rw,DeSimone:2007pa,DeSimone:2008ez,Tranberg:2008ae,
Hohenegger:2008zk,Anisimov:2008dz,Gagnon:2009zz,Garny:2009rv,Kiessig:2009cm}
for related work).

The Kadanoff--Baym formalism provides a powerful framework for studying
nonequilibrium processes within quantum field theory. However, it is
technically considerably more involved than the canonical Boltzmann approach.
For this reason, we apply it here to a simple toy model which we have already
used in \cite{Garny:2009rv} to investigate the vertex contribution.
The Lagrangian contains one complex and two real scalar fields:
\begin{align}
   \label{lagrangian}
   {\cal L} & = \frac12 \partial^\mu\psi_i\partial_\mu\psi_i
              - \frac12 M^2_i \psi_i\psi_i
              + \partial^\mu \bar{b}\partial_\mu b
              - m^2\bar{b}b \nonumber \\
            & - \frac{\lambda}{2!2!}(\bar{b}b)^2
              - \frac{g_i}{2!}\psi_i bb
              - \frac{g^*_i}{2!}\psi_i \bar{b}\bar{b}
              + {\cal L}_{rest}\,, i=1,2\,,
\end{align}
where $\bar{b}$ denotes the complex conjugate of $b$.
Here and in the following we assume summation over repeating 
indices, unless otherwise specified.
Despite its simplicity, the model incorporates all features relevant 
for leptogenesis. The real scalar fields imitate the (two lightest) 
heavy right-handed neutrinos, whereas the complex scalar field models 
the baryons. The $U(1)$ symmetry, which we use to define ``baryon'' number, 
is explicitly broken by the presence of the last two terms, just as the 
$B-L$ symmetry is explicitly broken by Majorana mass terms in 
phenomenological models. Thus the first Sakharov condition 
\cite{Sakharov:1967dj} is fulfilled. 
The couplings $g_i$ model the complex Yukawa couplings of the 
right-handed neutrinos to leptons and the Higgs. By rephasing the 
complex scalar field at least one of the couplings $g_i$ can be made real.
If ${\rm arg}(g_1)\neq {\rm arg}(g_2)$ the other one remains complex and 
there is  \textit{CP}-violation, as is required by the second Sakharov
condition. In vacuum the self-energy contribution to the \CP-violating
parameter is given by:
\begin{align}
   \label{epsilonclassic}
   \epsilon_i = -\frac{|g_j|^2}{16\pi}
   {\rm Im}\biggl(\frac{g_i g_j^*}{g_i^* g_j}\biggr)
   \frac{M_j^2-M_i^2}{(M_j^2 -M_i^2)^2+M_j^2 \Gamma_j^2}\,,
\end{align}
where $\Gamma_j$ is the decay width of the heavy scalar $\psi_j$.
 Note that the expression 
\eqref{epsilonclassic} has been obtained using the formalism developed in
\cite{Pilaftsis:2003gt}. The formalism employed in \cite{Anisimov:2005hr} leads
to a slightly different expression for the \CP-violating parameter (see Appendix
\ref{cpclassic}
for more details). The required deviation from thermal equilibrium is caused 
by the rapid expansion of the Universe. Thus the third Sakharov condition is 
fulfilled as well. Finally, the quartic self-interaction term in 
\eqref{lagrangian} plays the  role of the Yukawa and gauge interactions 
in established models -- it brings the ``baryons'' to equilibrium. The 
renormalizability of the theory requires the presence of some additional 
terms, which are accounted for by ${\cal L}_{rest}$. 

Here, we study  the self-energy contribution to the \CP-violating parameter in
the hierarchical and quasidegenerate cases using the
Schwinger--Keldysh/Ka\-danoff--Baym formalism as starting point.
To make the discussion less technical we give the  details of the 
calculation in the appendixes, whereas in the main body of the paper 
we sketch the derivation and present the results.
\begin{itemize}
\item[(i)]   As we argue in Sec.\,\ref{CPviol}, the formalism is free of the 
   double-counting problem typical for the canonical Boltzmann approach. 
   In other words the structure of the equations automatically ensures 
   that the asymmetry vanishes in thermal equilibrium and no need for the 
   real intermediate state (RIS) subtraction arises. This property has already
   been observed in case of the vertex contribution.
\item[(ii)]   The medium corrections to the \CP-violating parameter are only
linear
   in the particle number densities. That is, our result differs from that
   obtained previously in the framework of equilibrium thermal field
   theory by replacing the zero temperature propagators with finite
   temperature propagators in the matrix elements used in the Boltzmann
   equation.
\item[(iii)]   For scalars the medium effects always increase the \textit{CP}-violating 
   parameter, which in turn leads to an enhancement of the generated
   asymmetry.
\item[(iv)]   The canonical expression for the \CP-violating parameter is only 
   applicable in the hierarchical case even though it does not diverge
   in the limit of equal masses. For quasidegenerate masses
   one has to take into account quantum corrections to the effective masses
   and decay widths of the heavy particles in medium, which leads to a modified 
   expression for the \CP-violating parameter. 
\item[(v)] In the resonant regime quantum corrections also lead to an 
enhancement of the total \textit{in-medium} decay widths. This results 
in a faster decay of the heavy particles and can increase the importance 
of the washout process.
\item[(vi)] In the maximal
   resonant regime the Boltzmann picture breaks down and we argue that an 
   analysis in the full Kadanoff--Baym formalism is required.  
\end{itemize}
In Sec.\,\ref{Numerics} we present numerical solutions of the quantum-corrected 
Boltzmann equations, and discuss the quantitative impact of medium effects on
the final asymmetry within the toy model. Finally, in Sec.\,\ref{Summary}, we
summarize the results and present our conclusions.

% =============================================================================
\section{\label{kinEq}Kinetic equation}
% =============================================================================

The canonical approach to the calculation of the asymmetry 
generated at the epoch of leptogenesis is based on the use 
of Boltzmann equations. In the expanding Universe the 
Boltzmann equation for the toy-baryon distribution function 
$f_b$ can be written in the form \cite{Kolb:1990vq}
\begin{align}
   \label{BlzmnClassic} 
   p^\alpha {\cal D}_\alpha f_b =
   {\textstyle\frac12}\bigl[\Sigma_<(1+f_b)-f_b\,\Sigma_>\bigr]\,,
\end{align}
where all  functions are evaluated at the same point $(X,p)$
of the phase-space, ${\cal D}_\alpha$ is the covariant derivative 
and the quantities $\Sigma_\lessgtr$  correspond to the gain and loss 
terms. For decays into a pair of  toy-baryons they are given by
\begin{subequations}
   \label{Sigma_classic}
   \begin{align}
      \Sigma_<(X,p) \equiv & - \intg d\Pi^3_q d\Pi^3_k (2\pi)^4
                             \delta(k-q-p)\nonumber\\
                           & \times |{\cal M}|^2_{\psi_i\rightarrow bb}\,
                             f_{\psi_i}(X,k)[1+f_b(X,q)]\,,\\
      \Sigma_>(X,p) \equiv & - \intg d\Pi^3_q d\Pi^3_k (2\pi)^4
                             \delta(k-q-p) \nonumber\\
                           & \times |{\cal M}|^2_{bb\rightarrow \psi_i}\,
                             f_b(X,q)[1+f_{\psi_i}(X,k)]\,,
   \end{align}
\end{subequations}
where $d\Pi^3_p\equiv {d^3p}/[{(2\pi)^3}{2E}]$ is the invariant 
momentum-space volume element.
The analogous equations for the toy-antibaryon distribution function
$f_{\bar b}$ and the corresponding quantities $\bar \Sigma_\gtrless$ 
can be obtained from \eqref{BlzmnClassic} and \eqref{Sigma_classic} 
by replacing the subscript $b$ with $\bar b$. 

\begin{figure}[h!]
   \includegraphics[width=0.38\textwidth]{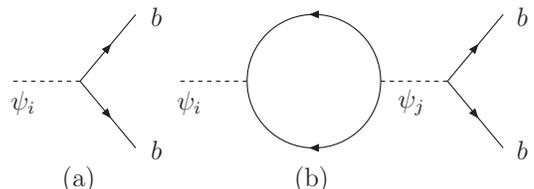}
   \caption{\label{interference}
      Tree-level and one-loop self-energy 
      diagrams of the decay process $\psi_i\rightarrow bb$.
   }
\end{figure}

At tree level, see Fig.\,\ref{interference} (a), the decay is 
\CP-conserving, that is $|{\cal M}|^2_{\psi_i \rightarrow bb}= 
|{\cal M}|^2_{bb \rightarrow \psi_i}\equiv |{\cal M}_0|^2$.
There are two distinct contributions to the 
\CP-violating parameter, the vertex \cite{Fukugita:1986hr,Garny:2009rv}  
and the self-energy \cite{Flanz:1994yx,PhysRevD.56.5431} ones.
The leading-order self-energy contribution is generated by the
one-loop self-energy  diagram depicted in Fig.\,\ref{interference} (b).
Only the diagram where the initial ($\psi_i$) and intermediate 
($\psi_j$) toy-Majoranas are different ($i\neq j$) contribute 
to the \CP-violating parameter. Summing the tree-level and the
``off-diagonal'' one-loop self-energy amplitudes one obtains
$|{\cal M}|^2_{\psi_i \rightarrow bb}=|{\cal M}_0|^2(1+\epsilon_i)$
and
$|{\cal M}|^2_{bb \rightarrow \psi_i}=|{\cal M}_0|^2(1-\epsilon_i)$.
In the corresponding expressions for the ``antibaryons'' the signs in 
front of $\epsilon_i$ are reversed. If  these amplitudes are now 
substituted into $\Sigma_\gtrless$ and $\bar \Sigma_\gtrless$ one 
finds that a nonzero asymmetry is generated even in thermal 
equilibrium, which is inconsistent with the \textit{CPT} symmetry.
This is a manifestation of the so-called double-counting 
problem typical for the canonical Boltzmann formalism. 
Let us briefly explain what that means. An inverse decay 
$bb\rightarrow \psi_i$ immediately followed by a decay 
$\psi_i \rightarrow \bar b \bar b$ is equivalent to the two-body 
scattering process $bb\rightarrow \psi_i \rightarrow \bar b \bar b$
where the intermediate toy-Majorana is on the mass shell. That 
is, the same contribution is taken into account twice: once 
when  the decay and inverse decay processes are considered and 
once when  the two-body scattering processes are considered. This
problem is usually solved by subtracting the contribution of the 
on-shell intermediate state to the scattering amplitude  
\cite{PhysRevD.56.5431,Pilaftsis:2003gt}. Roughly speaking, 
after the subtraction the amplitude of the inverse decay process
$bb \rightarrow \psi_i$ also becomes proportional to  $(1+\epsilon_i)$,
which ensures that in thermal equilibrium no asymmetry is produced
 due to detailed balance.

In the canonical \textit{bottom-up} approach, which has been outlined above, one
uses elements of the \textit{S}-ma\-trix (in-out formalism) to calculate the
functions $\Sigma_\gtrless$. In contrast to that, in the \textit{top-down} approach based
on the Schwi\-nger--Keldysh/Kadanoff--Baym
\cite{Schwinger:1960qe,Keldysh:1964ud,
Bakshi:1963ab,Danielewicz:1982kk,Chou:1984es,Calzetta:1986cq,Ivanov:1999tj,
Knoll:2001jx,Blaizot:2001nr,Berges:2002wt,Berges:2005md,Weinstock:2005jw,
Carrington:2004tm,FillionGourdeau:2006hi} formalism, the functions
$\Sigma_\gtrless$ can  be identified with self-energies and are derived,
using nonequilibrium field theory techniques (see e.g.~\cite{Berges:2004yj,
Carrington:2004tm}), from the two-particle-irreducible (2PI) effective action%
\footnote{%
   Note that there exists a straightforward generalization to a hierarchy
   of so-called nPI effective actions~\cite{Berges:2004pu}. In general, for
   systems far from equilibrium and for strong couplings the 2-body decay
   is best described using 3PI. However, for leptogenesis, 2PI is
   sufficient since the CP-violation is dominated by the leading loop
   diagrams.
}%
\cite{PhysRevD.10.2428} formulated on the closed real-time path (in-in
formalism).
\begin{figure}[h!]
   \includegraphics[width=0.48\textwidth]{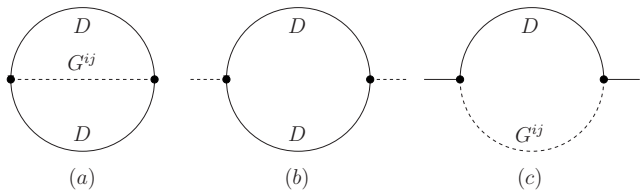}
   \caption{\label{diagrams}
      Two-loop contribution to the 2PI effective action and the
      corresponding contributions to the self-energies of the real
      and complex fields.
   }
\end{figure}  
A two-loop contribution to the effective action and the corresponding 
contribution to the self-energy of the ``baryons'' are presented in 
Fig.\,\ref{diagrams} (a) and Fig.\,\ref{diagrams} (c) respectively. 
Wigner-transforming the self-energy we obtain \cite{Garny:2009rv} 
\begin{subequations}
   \label{SigmasWT}
   \begin{align}
      \label{SigmaWT}
      \Sigma_\gtrless(X,p)      = & -\intg d\Pi_{k}d\Pi_{q}(2\pi)^4
                                    \delta^g(k-q-p)\nonumber\\
                                  & \times \, g^*_i g_j\,
                                    G^{ij}_\gtrless(X,k)
                                    D_\lessgtr(X,q)\,,\\
      \label{SigmaWTBar}
      \bar \Sigma_\gtrless(X,p) = & -\intg d\Pi_{k}d\Pi_{q}(2\pi)^4
                                    \delta^g(k-q-p)\nonumber\\
                                  & \times \, g_i g^*_j\,
                                    G^{ij}_\gtrless(X,k)
                                    \bar D_\lessgtr(X,q)\,,
   \end{align}
\end{subequations}
where $d\Pi_p\equiv {d^4p}/{(2\pi)^4}$, $\delta^g(p)\equiv \sqrt{-g}_X\,
\delta(p)$ is the covariant 
generalization of the Dirac $\delta$-function and $G^{ij}_\gtrless$ and
$D_\gtrless$
are the Wightman propagators of the real and complex fields respectively.
In the Boltzmann-limit the latter are related to the distribution function by 
\begin{align}
   \label{Dgtrless}
   D_<=f_b\,D_\rho,\quad D_>=(1+f_b)\,D_\rho\,,
\end{align}
where in the quasiparticle approximation (see \cite{Garny:2009rv} for more
details) the spectral function reads
\begin{align}
   \label{DrhoQP}
   {D}_{\rho}(X,p)=2\pi\,\sign(p_0) \,
   \delta\left(g^{\mu\nu} p_\mu p_\nu-m^2\right)\,.
\end{align}
In the same approximation the Wightman propagators of the toy-antibaryons, 
$\bar D_\gtrless$, can be obtained from \eqref{Dgtrless} by replacing $f_b$ 
with $f_{\bar b}$. 

If the real scalar fields would not mix, i.e.~if the off-diagonal
components of the Wightman propagator were equal to zero, we 
could also write analogous expressions for the diagonal components:
\begin{align}
   \label{Giigtrless}
   G^{ii}_< = f_{\psi_i}\,G^{ii}_\rho\,,\quad
   G^{ii}_> = (1+f_{\psi_i})\,G^{ii}_\rho\,,
\end{align}
with the spectral function given by
\begin{align}
   \label{GrhoQP}
   G^{ii}_{\rho}(X,p)=2\pi\,\sign(p_0) \, 
   \delta\left(g^{\mu\nu} p_\mu p_\nu-M_i^2\right)\,.
\end{align}
Because of the presence of the Dirac $\delta$-function in \eqref{DrhoQP}
and \eqref{GrhoQP} the integrations over $k^0$ and $q^0$ could then 
be perfor\-med trivially and we would recover \eqref{Sigma_classic} 
in the \CP-\textit{con\-ser\-ving} limit. 

A nonzero toy-baryon asymmetry can be generated only if 
$\Sigma_\gtrless\neq \bar\Sigma_\gtrless$. Comparing \eqref{SigmaWT} 
and \eqref{SigmaWTBar} we see that if initially the system is symmetric,
i.e. $f_b=f_{\bar b}$,  this amounts to the requirement that the Hermitian 
matrix $\hat G_\gtrless$ has complex off-diagonal components. If  the 
off-diagonal components peak on the mass shell of the quasiparticle species,
i.e.~if they can be represented in the form
\begin{align}
   \label{GgtrlessDecomp}
   G^{ij}_\gtrless=\varepsilon_i \, G^{ii}_\gtrless+\varepsilon^*_j\,
   G^{jj}_\gtrless\quad (i\neq j), 
\end{align}
then the generation of the asymmetry can be analyzed in terms of  
\CP-violating parameters. Substituting \eqref{GgtrlessDecomp} into 
\eqref{SigmasWT} we find
\begin{align}
   \label{EpsilonExpected}
   \epsilon_i = -2\,\Im\left(g_j/g_i\right) \Im \, \varepsilon_i\,.
\end{align}
To calculate the decomposition coefficients $\varepsilon_i$ we 
use the nonequilibrium formulation of the Schwinger-Dyson 
equation, which is discussed in the following section.

% =============================================================================
\section{\label{CPviol}CP-violating parameter}
% =============================================================================

Just like in the canonical analysis \cite{PhysRevD.56.5431,Pilaftsis:2003gt}
the starting point of our analysis is the Schwinger--Dyson equation 
\begin{align}
   \label{SchwingerDyson}
   [G^{-1}]^{ij}(x,y) = [\mathscr{G}^{-1}]^{ij}(x,y)-\Pi^{ij}(x,y)\,,
\end{align}
where $G^{ij}$ is the full dressed  propagator  of the 
``heavy neutrinos'', $\mathscr{G}^{ij}$ is the diagonal 
propagator of the free fields and $\Pi^{ij}$ is the self-energy. 
In the two-loop approximation, see Fig.\,\ref{diagrams}, 
the self-energy is given by 
\begin{align}
   \label{FullPi}
   \Pi^{ij}(x,y) = - {\textstyle\frac12}g_i g^*_j D^2(x,y)
                   - {\textstyle\frac12}g^*_i g_j \bar D^2(x,y)\,,
\end{align}
where $D(x,y)$ is the full propagator of the complex scalar field and
$\bar D(x,y)\equiv D(y,x)$. Note that the arguments of the two-point functions
and self-energy are defined on the positive and negative branches of the
Schwinger--Keldysh closed real-time contour $\cal C$ \cite{Schwinger:1960qe,
Keldysh:1964ud,Bakshi:1963ab,Danielewicz:1982kk,Chou:1984es,Calzetta:1986cq}
shown in Fig.\,\ref{contour}.

If we multiply \eqref{SchwingerDyson} by $G$ from the 
right, decompose the full propagator and the self-energy 
into  statistical and spectral components and  
integrate over the contour, we  obtain a system of 
so-called Kadanoff--Baym equations for the statistical propagator
and spectral function (see Appendix \ref{real}).
Within leptogenesis, the light particle species are
usually assumed to be very close to kinetic equilibrium. 
Translated to the toy model, this means that the toy-baryons 
described by the propagators $D(x,y)$ and $\bar D(x,y)$ are close
to equilibrium. In this case, one can approximate the full 
Kadanoff--Baym equations by quantum kinetic equations, and 
find approximate analytic solutions of these equations.
Using these solutions, it is possible to explicitly obtain the 
decomposition coefficients $\varepsilon_i$, which then
yield the corresponding \CP-violating parameters $\epsilon_i$ 
as described above. This approach is pursued in Appendix \ref{real}. 
In this section we will use an alternative approach, which essentially 
relies on the same assumptions, but is somewhat more elegant.

In the following, we use a compact matrix notation for 
Eq.\,(\ref{SchwingerDyson}), where we denote the matrices by 
a hat, e.g.~$\hat G \equiv (G^{ij})$.
Let us split the self-energy matrix  $\hat \Pi$ into 
the diagonal $\hat \varPi$ and off-diagonal $\hat \varPi^{'}$ 
components and introduce a diagonal propagator $\hat {\cal G}$ 
defined by the equation
\begin{align}
   \label{DiagonalEq}
   \hat {\cal G}^{-1}(x,y)=\hat{\mathscr{G}}^{-1}(x,y)-\hat \varPi(x,y)\,.
\end{align}
Subtracting \eqref{DiagonalEq} from \eqref{SchwingerDyson} we find
\begin{align}
   \label{OffDiagDiagonalEq}
   \hat G^{-1}(x,y) = \hat {\cal G}^{-1}(x,y)-\hat \varPi^{'}(x,y)\,.
\end{align}
Multiplying \eqref{OffDiagDiagonalEq} by $\hat G$ from the left, by 
$\hat {\cal G}$ from the right and integrating over the contour 
$\cal C$ we obtain a formal solution for the full nonequilibrium 
propagator:%
\footnote{%
   Here, we implicitly assume that the correlators are diagonal 
   at the initial time of the closed time path evolution.
   Since we consider the kinetic limit later on, for which the 
   initial time is formally sent to negative infinity, the results
   are not affected.
}
\begin{align}
   \label{FormalSolution}
   \hat G(x,y) & = \hat {\cal G}(x,y)\nonumber\\
               & + {\textstyle \iint\limits_{{\cal C}\, {\cal C}} }
                   \mathscr{D}^4u \,\mathscr{D}^4v \,\, \hat G(x,u)
                   \hat \varPi^{'}(u,v) \hat {\cal G}(v,y)\,.
\end{align}
The invariant volume element, $\dg u\equiv \sqrt{-g} d^4 u\,$, where 
$\quad g\equiv \det g_{\mu\nu}\,$, ensures that \eqref{FormalSolution} can be
applied to the analysis of out-of-equilibrium dynamics not only in Minkowski, 
but also in a general curved space-time \cite{Hohenegger:2008zk}.
Using  the decomposition
\begin{align}
   \label{Gdecompos}
   \hat G(x,y) =  \hat G_F(x,y) - {\textstyle\frac{i}{2}}\,
   \sign_{\cal C}(x^0-y^0) \hat G_\rho(x,y)\,,
\end{align}
and an analogous relation for the self-energy 
we can split \eqref{FormalSolution} into spectral and statistical
components. These are related to the Wightman propagators by
\begin{align}
   \label{GgtrlessDef}
   \hat G_\gtrless(x,y) = \hat G_F(x,y)\mp{\textstyle\frac{i}2}\,
   \hat G_\rho(x,y)\,.
\end{align}
Because of the $\sign_{\cal C}$ function in the decomposition
of the two-point function and self-energy the integrals over 
the closed-time-path contour in \eqref{FormalSolution} reduce
to integrals over parts of the $uv$ plane; see Appendix 
\ref{ContourIntegration}. Introducing the 
retarded and advanced propagators,
\begin{subequations}
   \label{RetAdvDef}
   \begin{align}
      \hat G_R(x,y) & \equiv \theta(x^0-y^0)\hat G_\rho(x,y)\,,\\
      \hat G_A(x,y) & \equiv -\theta(y^0-x^0)\hat G_\rho(x,y)\,,
   \end{align}
\end{subequations}
and also the retarded and advanced self-energies we can 
represent the resulting expressions as integrals over
the whole $uv$ plane. Finally, building the linear 
combinations \eqref{GgtrlessDef} we find for the Wightman propagators
\begin{align}
   \label{GgtrlessFormalSol}
     \hat G_\gtrless(x,y)= \hat {\cal G}_\gtrless(x,&\,y)
     -{\textstyle\iint} \mathscr{D}^4u \mathscr{D}^4v\, \theta(u^0) \,
     \theta(v^0)\nonumber\\
   \times \bigl[& \hat G_R(x,u) \hat \varPi^{'}_\gtrless(u,v)
     \hat {\cal G}_A(v,y)\nonumber\\
   + & \hat G_\gtrless(x,u)\hat \varPi^{'}_A(u,v)\hat {\cal G}_A(v,y)
     \nonumber\\
   + & \hat G_R(x,u)\hat \varPi^{'}_R(u,v)\hat {\cal G}_\gtrless(v,y)\bigr]\,.
\end{align}
Using \eqref{GgtrlessDef} and \eqref{RetAdvDef} we can also derive formal
solutions for the retarded and advanced pro\-pagators from
\eqref{GgtrlessFormalSol}. They read
\begin{align}
   \label{RetAdvFormalSol}
   \hat G_{R(A)}(x,y) & = \hat {\cal G}_{R(A)}(x,y)-{\textstyle\iint}
                        \mathscr{D}^4u \mathscr{D}^4v\, \theta(u^0) \,
                        \theta(v^0)\nonumber\\
                      & \times \hat G_{R(A)}(x,u)
                        \hat\varPi^{'}_{R(A)}(u,v)
                        \hat {\cal G}_{R(A)}(v,y)\,.
\end{align}
In thermal equilibrium the two-point functions and self-energies 
in \eqref{GgtrlessFormalSol} and \eqref{RetAdvFormalSol} depend only 
on the relative coordinate, $s\equiv x-y$, and are independent of 
the center coordinate, $X\equiv \frac12(x+y)$, i.e.~are 
translationally invariant. As discussed above, we expect that
the deviations from equilibrium are  moderate. In this case,  
one can perform a gradient expansion of the two-point 
functions and the self-energies in the vicinity of $X$ keeping only the leading
terms. Performing the  Wigner transformation, i.e.~the Fourier transformation
with respect to the relative coordinate  (see Appendix\,\ref{real}) 
we trade the relative coordinate $s$ for a coordinate $p$ in  
momentum space. Effectively, the Wigner transformation replaces
the coordinate-space arguments $(x,y)$ of each two-point function 
in \eqref{GgtrlessFormalSol} and \eqref{RetAdvFormalSol} by 
the phase-space  coordinates $(X,p)$ and ``removes'' the 
double integration; see Appendix \ref{ContourIntegration}.
Combining the Wigner transforms of \eqref{GgtrlessFormalSol} 
and \eqref{RetAdvFormalSol}  we find  for the full Wightman propagators
\begin{align}
   \label{GgtrlessSolWT}
   \hspace{-1mm}
   \hat G_\gtrless = \frac{\bigl[\hat I -\hat {\cal G}_R \hat \varPi^{'}_R\bigr]
   \bigl[\hat {\cal G}_\gtrless-\hat {\cal G}_R \hat \varPi^{'}_\gtrless \hat
{\cal G}_A\bigr]
   \bigl[\hat I - \hat \varPi^{'}_A \hat {\cal G}_A \bigr]}{
   \det\bigl[\hat I -\hat {\cal G}_R \hat \varPi^{'}_R\bigr] 
   \det \bigl[\hat I - \hat \varPi^{'}_A \hat {\cal G}_A \bigr]}\,,
\end{align}
where all the functions are evaluated at the same point $(X,p)$ of the 
phase-space. The term proportional to $\hat \varPi^{'}_\gtrless$ describes
scattering \cite{PhysRevB.52.14615,PhysRevC.46.1687,PhysRevC.48.1034,
PhysRevC.64.024613,CondMatPhys2006_9_473,JPhys2006_35_110,Weinstock:2005jw,
Carrington:2004tm,FillionGourdeau:2006hi} and is irrelevant 
for us at the moment. Contracting the decay term with the product of the 
couplings in \eqref{SigmaWT}  and using that, by construction, the propagators
${\cal G}$ are diagonal and $\varPi'$ off-diagonal,  we find
\begin{align}
   \label{Contraction}
   g^*_i g_j G^{ij}_\gtrless = 
   \frac{{\textstyle \sum_i}\, {\cal
G}_\gtrless^{ii}|g_i|^2\bigl[1+\Delta^i_b\bigr]}{
   \det\bigl[\hat I -\hat {\cal G}_R \hat \varPi^{'}_R\bigr] 
   \det \bigl[\hat I - \hat \varPi^{'}_A \hat {\cal G}_A \bigr]}\,,
\end{align}
where 
\begin{align}
   \label{DeltaB}
   |g_i|^2 \Delta^i_b\equiv |g_j|^2 \Pi_A^{ij} \Pi_R^{ji}
   {\cal G}_R^{jj}{\cal G}_A^{jj} -
   2\Re\bigl[g_i g_j^* {\cal G}_R^{jj}\Pi_R^{ji}\bigr]\,.
\end{align}
In \eqref{DeltaB} and all the equations below we implicitly assume 
summation over all $j\neq i$.

A very important feature of \eqref{Contraction} is that the loop corrections
$\Delta^i_b$   are the same for both the gain and loss terms (i.e. for the $>$
and $<$ components), respectively. To obtain an equivalent result in the
canonical approach one needs to apply the real intermediate state subtraction
procedure \cite{PhysRevD.56.5431,Pilaftsis:2003gt}. This means that, here, the
structure of the equations automatically ensures that no asymmetry is generated
in thermal equilibrium. Stated differently, the formalism is \emph{free of the
double-counting problem} and no need for RIS subtraction arises. This conclusion
is in accordance with the corresponding result for the vertex contribution to
the \CP-violating parameter~\cite{Garny:2009rv}.

Contracting the decay term with the product of the couplings in
\eqref{SigmaWTBar} we obtain an expression similar to \eqref{Contraction} but
now with
\begin{align}
   \label{DeltaBBar}
   |g_i|^2 \Delta^i_{\bar b}\equiv |g_j|^2 \Pi_A^{ij} \Pi_R^{ji} 
   {\cal G}_R^{jj}{\cal G}_A^{jj}
   - 2\Re\bigl[g_i^* g_j {\cal G}_R^{jj}\Pi_R^{ji}\bigr] \,.
\end{align}
From Eq.\,(\ref{SigmasWT}), we see that the rates for the decays
$\psi_i\rightarrow bb$ and $\psi_i\rightarrow \bar b\bar b$ differ from each
other only if $\Delta^i_b\not= \Delta^i_{\bar b}$. The corresponding
\CP-violating parameter is  given by
\begin{equation}
   \label{EpsilonDelta}
   \epsilon_i = \frac{ [1+\Delta_b^i] - [1+\Delta_{\bar b}^i] }
   { [1+\Delta_b^i] + [1+\Delta_{\bar b}^i] } \;.
\end{equation}
The first terms in \eqref{DeltaB} and \eqref{DeltaBBar} are equal and cancel
out in the difference of $\Delta^i_b$ and $\Delta^i_{\bar b}$. Therefore, in
agreement with \eqref{EpsilonExpected}, the  \CP-violating parameter can be
expressed as 
\begin{equation}
   \label{EpsilonDef}
   \epsilon_i =
   \frac{-2\,\Im\bigl(g_j/g_i\bigr)\, \Im\bigl({\cal G}_R^{jj} \Pi_R^{ji}\bigr)
}
   { 1 + \frac12\left( \Delta_b^i + \Delta_{\bar b}^i \right) } \,.
\end{equation}

Equation \eqref{Contraction} also provides us with an expression for the total
in-medium decay widths of the heavy particles:
\begin{align}
   \label{GammaInMedium}
   \Gamma_i^{med}=\Gamma_i\times
   \frac{1+{\textstyle\frac12}(\Delta^i_b+\Delta^i_{\bar b})}{
   \det\bigl[\hat I -\hat {\cal G}_R \hat \varPi^{'}_R\bigr] 
   \det \bigl[\hat I - \hat \varPi^{'}_A \hat {\cal G}_A \bigr]}\,,
\end{align}
where $\Gamma_i=|g_i|^2/(16\pi M_i)$ are the corresponding tree-level decay
widths in vacuum.

The \CP-violating parameter \eqref{EpsilonDef} is a function of the space-time 
coordinate $X$ and four-momentum $p$. Note that in general 
$p$ does not need to be on-shell. The on-shell condition is 
determined by the diagonal components of the full propagator.
The \CP-violating parameter carries information about the 
off-diagonal components, compare Eq.\,\eqref{GgtrlessDecomp}. Therefore, the
conventional, ``on-shell'', interpretation of \eqref{EpsilonDef} is only 
applicable if the off-diagonal components of the full propagator 
peak at the same values of the momentum as the diagonal ones.  
Furthermore, if we want to use the Boltzmann equations 
for $\hat {\cal G}$ to calculate the asymmetry, then the 
mass spectra of the diagonal ($\hat {\cal G}$) and 
full  ($\hat G$) propagators must be sufficiently close.

The value of the \CP-violating parameter depends on the masses $M_i$ 
of the heavy species  and the coupling constants $g_i$ or, alternatively,
the decay widths $\Gamma_i$. It is useful to discriminate between three 
cases, namely
\begin{align}
      M_1^2 & \ll M_2^2,  M_{j}^2 \gg M_j\Gamma_j ~~~~
              \mbox{(strongly hierarchical case)} \,, \nonumber\\
      M_i^2 & \gg  |\Delta M_{ij}^2|  \gg M_j\Gamma_j ~~~~~
              \mbox{(resonant case)}\,, \\
      M_i^2 & \gg  |\Delta M_{ij}^2|  \lesssim M_j\Gamma_j ~~~~~~
              \mbox{(maximal resonant case) \,,}\nonumber
\end{align}
where $\Delta M_{ij}^2 \equiv M_i^2 - M_j^2$. As we will argue in the 
following, the generation of the asymmetry can be studied approximately 
in the Boltzmann picture using the above \CP-violating parameter in the 
\textit{strongly hierarchical} and \textit{resonant} cases, but not in the
\textit{maximal resonant} case.

\subsection{Strongly hierarchical case}

In this subsection, we derive the \CP-violating parameter in the strongly
hierarchical limit, $M_1 \ll M_2$. As we shall see later, in the hierarchical
case the denominator in Eq.\,\eqref{EpsilonDef} is close to unity. 
The same is true for the denominator of \eqref{GammaInMedium}. 
Therefore the in-medium decay widths $\Gamma_i^{med}$ coincide with the
tree-level ones. For the \CP-violating parameter we obtain
\begin{align}
   \label{EpsilonDefHierarchical}
   \epsilon_i = 
   -2\,\Im\bigl(g_j/g_i\bigr)\,\Im\bigl({\cal G}_R^{jj} \Pi_R^{ji}\bigr)\,.
\end{align}
It only remains to calculate the last term in \eqref{EpsilonDefHierarchical}.
Let us first analyze the self-energy $\hat \Pi_R$.
In a ``baryonically'' symmetric configuration the particles and antiparticles 
are interchangeable, so that $\bar D(x,y)=D(x,y)$ (see Appendix \ref{SymmConf}
for a proof). In the following, we will use that the baryon asymmetry is small,
i.e. $D(x,y) \simeq \bar D(x,y)$. In this case the two terms of \eqref{FullPi}
combine to a symmetric matrix which is proportional to $(g_i^*g_j+g_i g_j^*)$. 
To obtain $\hat \Pi_R(X,p)$ we decompose this matrix into a statistical 
and spectral part, multiply the spectral part with $\theta(x^0-y^0)$ and
perform the Wigner transformation. The result reads
\begin{align}
   \label{PiRSymmetric}
   \Pi^{ij}_{R}(X,p) = & -(g_i^*g_j+g_i g_j^*)
                         {\textstyle \int} d\Pi_k d\Pi_q (2\pi)^4
                         \delta(p-k-q)\nonumber\\
                       & \times D^{(s)}_F(X,k)D^{(s)}_{R}(X,q)\,,
\end{align}
where the superscript `s' refers to the ``baryonically'' symmetric 
configuration and  $D_F$ is the statistical propagator of the complex 
field. In the quasiparticle approximation $D_F=(\frac12+f_b)D_\rho$.
The retarded and advanced propagator of the complex scalar field can be 
represented in the form $D_{R(A)}=D_h \pm \frac{i}{2}D_\rho$, where in the 
quasiparticle approximation $D_\rho$ is given by \eqref{DrhoQP} 
and $D_h=-\frac{{\cal P}}{p^2-m^2}$; see Eq.\,\eqref{D_h}. 
Consequently, in the symmetric configuration the retarded self-energy 
is given by a linear combination of two real-valued symmetric matrices: 
\begin{align}
   \label{PiRADecomposition}
   \Pi^{ij}_{R}=\Pi^{ij}_{h} & + {\textstyle\frac{i}2}\Pi^{ij}_{\rho}
   = {\textstyle\frac{-1}{16\pi}}(g_i^*g_j+g_i g_j^*)
   (L_h+{\textstyle\frac{i}2}L_\rho),
\end{align}
where, to shorten the notation, we have introduced 
\begin{align}
   \label{LRhoDef}
   L_{h(\rho)}(X,p)\equiv 16\pi\,{\textstyle \int}  d\Pi_k & d\Pi_q   (2\pi)^4  
   \delta(p-k-q)\nonumber\\ &\times D^{(s)}_F(X,k)D^{(s)}_{h(\rho)}(X,q)\,.
\end{align}
For a homogeneous and isotropic system the one-particle distribution
functions depend only on the Lorentz-invariant product $ku$, where 
$k$ is the particles' momentum and $u$ is the (constant) four-velocity of the
medium with respect to the chosen frame of reference. Using this 
property, in Appendix \ref{OneLoopInt} we evaluate $L_\rho$ on the mass shell 
of the $i$'th quasiparticle:
\begin{align}
   \label{LRho}
   L_\rho(X,p) = r {\textstyle \int\frac{d\Omega}{4\pi}}
   [1+f_b(E_p)+f_{\bar b}(E_p)]\,,
\end{align}
where $E_p \equiv {\textstyle\frac12} [E_i+r|\vec{p}|\cos\Theta]$. Here,
$E_i\equiv(M_i^2+\vec{p}^2)^\frac12$ and $\vec{p}$ are the components of the 
four-momentum $p$ of the  heavy  scalar in the rest-frame of the medium, 
and $r\equiv (1-4m^2/M_i^2)^\frac12$.
For massless toy-baryons in vacuum $L_\rho^{vac}|_{m=0}=1$. Taking a nonzero 
toy-baryon mass $m$ into account yields $L_\rho^{vac} = r$.
In medium (symmetric case), we find using Eq.\,\eqref{LRho}
\begin{align}\label{LRhoMed}
   \frac{L_\rho(X,p)}{L_\rho^{vac}} = & 1 + {  \frac{2}{r\left|\vec{p}\right|}} 
   \int_{E_{min}/2}^{E_{max}/2} \!\!\!\! f_b(E) dE \, ,
\end{align}
where $E_{\it max}=E_i+r|\vec{p}|$ and $E_{\it min}=E_i-r|\vec{p}|$ are 
the largest and smallest kinematically allowed energies of the light 
scalars produced in the decay $\psi_i\rightarrow bb$. 
 
As we show in Appendix \ref{real}, analogously to \eqref{PiRADecomposition} 
the Wigner transforms of the retarded and advanced propagators can 
be represented as  linear combinations of two Hermitian matrices. 
Applied to the matrix  $\hat {\cal G}$ this implies that 
it splits into two real-valued diagonal matrices:
\begin{align}
   \label{GRADecomposition}
   {\cal G}^{jj}_{R(A)} =  {\cal G}^{jj}_h \pm
   {\textstyle \frac{i}2}  {\cal G}^{jj}_\rho\,.
\end{align}
In the hierarchical case we can neglect the finite decay widths of the 
quasiparticle species and approximate ${\cal G}_\rho^{jj}$ by 
\eqref{GrhoQP}. It diverges on the mass shell of the $j$'th 
quasiparticle and is zero everywhere else. Because of the presence 
of $G_\gtrless^{ii}$ in \eqref{Contraction}, the \CP-violating 
parameter \eqref{EpsilonDefHierarchical} must be evaluated on the mass shell 
of the $i$'th quasiparticle. Since the product of $G_\gtrless^{ii}$
and ${\cal G}_\rho^{jj}$ vanishes in the quasiparticle approximation, 
we conclude that only the ${\cal G}_h$ term contributes. It is given 
by \cite{Hohenegger:2008zk}
\begin{align}
   \label{GhReal}
   {\cal G}^{ii}_h= - \frac{p^2-M_i^2-\Pi_h^{ii}}{(p^2-M_i^2-\Pi_h^{ii})^2+
   {\textstyle\frac14}(\Pi_\rho^{ii})^2}\,.
\end{align}
In vacuum the tree-level decay width of the heavy species is given 
by $\Gamma_i=|g_i|^2/(16\pi M_i)$. Therefore, we can rewrite the 
diagonal components of the spectral self-energy in the form 
$\Pi_\rho^{ii}=-2 M_i \Gamma_i L_\rho$\,. 

In the hierarchical case $|M_i^2-M_j^2|\gg |\Pi_h^{ii}-\Pi_h^{jj}|$. 
Substituting \eqref{PiRADecomposition} and \eqref{GRADecomposition}
into \eqref{EpsilonDefHierarchical}, and evaluating the
momentum on-shell, $p^2=M_i^2$,  we then obtain for the \CP-violating 
parameter\footnote{Note that, by setting $p^2=M_i^2$, we neglect
possible off-shell effects (see Appendix~\ref{cpclassic}). This is a good
approximation if $|M_i-M_j|\gg \Gamma_i,\Gamma_j$, which is the case in the
strongly hierarchical case.}
\begin{align}
   \label{EpsilonHierarchical}   
   \epsilon_i &= -\Im(g_j/g_i)\,{\cal G}_h^{jj}\, \Pi_\rho^{ji}\\
              &= -\frac{|g_j|^2}{16\pi}\Im\biggl(\frac{g_i g_j^*}{g_i^*
g_j}\biggr)
                 \frac{M_j^2-M_i^2}{(M_j^2-M_i^2)^2+(M_j \Gamma_j)^2\, L_\rho^2}
                 \cdot L_\rho \,. \nonumber
\end{align}
For massless toy-baryons in vacuum $L_\rho=1$ and we recover the 
classical result \eqref{epsilonclassic}.
In the strongly hierarchical limit, one can neglect the decay width $\Gamma_j$
in the denominator of \eqref{EpsilonHierarchical}. In this case,
\begin{equation}
   \epsilon_i/\epsilon_i^{vac} = L_\rho(X,p)/L_\rho^{vac} \,.
\end{equation}

This result is identical to the one for the vertex contribution \cite{Garny:2009rv}.
The \textit{CP}-violating parameter is given by a sum of  vacuum and medium 
contributions. From Eq.\,\eqref{LRhoMed}, we find that the medium
contributions are proportional to the one-particle distribution function,
which is  positive. Hence, for the scalar toy model and the strongly
hierarchical limit considered here, the self-energy contribution to the
\textit{CP}-violating parameter is always \textit{enhanced} by the medium
effects.

Note that the medium contribution depends only on the distribution function 
of the toy-baryons and is independent of that of the toy-Majoranas.
Since we expect the light scalars to be close to kinetic equilibrium at all 
times, it is instructive to estimate the size of the corrections in thermal 
equilibrium. In the hierarchical limit the asymmetry is predominantly generated 
by the decay of the lighter toy-Majorana. Inserting a Bose--Einstein
distribution (BE) or a Maxwell--Boltzmann (MB) distribution function we obtain for the 
ratio of the corresponding \CP-violating parameter and its vacuum value:
\begin{equation*}
\label{epsilon_hierarchical_BE}
   \frac{ \epsilon_1(|\vec{p}|)}{ \epsilon_1^{\it vac} } \, =
   \, 1 + \frac{2T}{r|\vec{p}|} \times \left\{
   \begin{array}{ll}
         \displaystyle \ln \left( \frac{ 1 - \exp\left( - \frac{E_{\it max}
-2\mu }{ 2T } \right) }
         { 1 - \exp\left( - \frac{E_{\it min} - 2\mu }{ 2T } \right) } \right)
      &  \mbox{BE} \; , \\[4.5ex]
         \displaystyle e^{ - \frac{E_{\it min} -2\mu }{ 2T } }
         - e^{ - \frac{E_{\it max} -2\mu }{ 2T } }
      &  \mbox{MB} \;.
   \end{array}\right.
\end{equation*}

\begin{figure}[ht!]
   \includegraphics[width=\columnwidth]{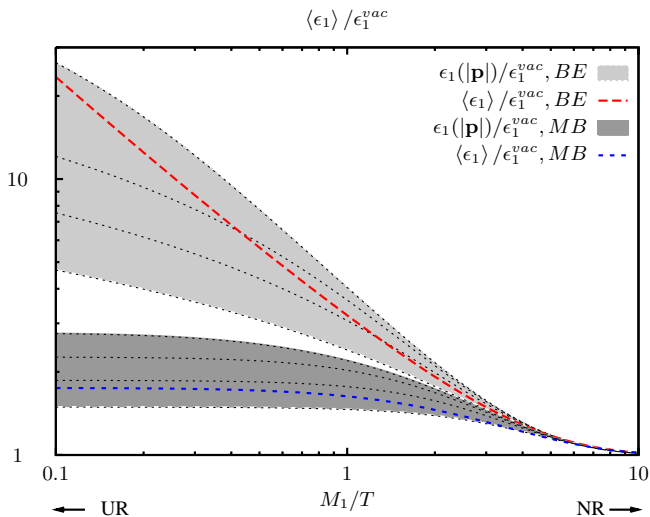}
   \caption{\label{thermalEpsilonAveraged}%
      Effective self-energy \CP-violating parameter $\epsilon_1$ in medium
      obtained from the Kadanoff--Baym formalism. The shaded areas correspond
      to the range $0.25 \le |\vec{p}|/T \le 4$ of momenta $|\vec{p}|$ of the
      decaying particle $\psi_1$ with respect to the rest-frame of the medium.
      Here we assumed a thermal Bose-Einstein (BE) or Maxwell-Boltzmann (MB)
      distribution for $b$/$\bar b$ with vanishing chemical potential. In the
low-temperature limit (NR), the vacuum value is
      approached. In the high-temperature limit (UR), the \CP-violating
      parameter is enhanced within the toy model. We also show the thermally
averaged 
      \CP-violating parameter $\langle\epsilon_1\rangle$ for the BE 
      (red long-dashed line) and MB (blue dashed line) cases.
   }
\end{figure}
The temperature- and momentum dependence of the medium correction in the range
of typical momenta $|\vec{p}| \sim T$ is shown in the shaded areas in
Fig.\,\ref{thermalEpsilonAveraged} for the BE and MB cases, respectively. We
also  show the \CP-violating parameter $\langle\epsilon_1\rangle$ obtained by
averaging Eq.\,(\ref{EpsilonHierarchical}) over the momentum $|\vec{p}|$. As
expected, $\langle\epsilon_1\rangle \sim \epsilon_1(|\vec{p}|\sim T)$.

As we have argued above, the conventional interpretation of the \CP-violating
parameter is only possible if the off-diagonal components of the full
propagator peak at the same values of the momentum as the diagonal ones. 
\begin{figure}[h!]
   \includegraphics[width=0.9\columnwidth]{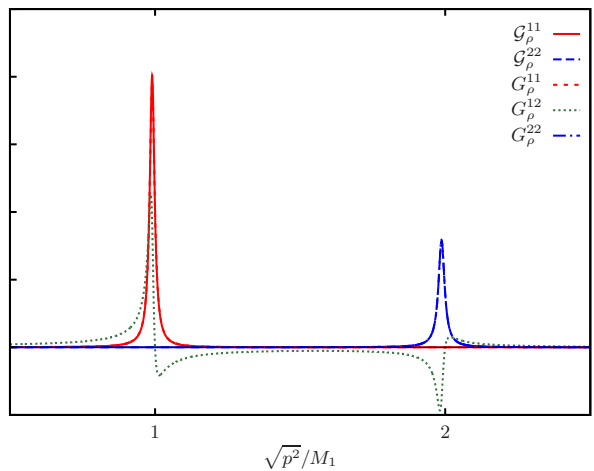} 
   \caption{\label{GrhoHierarchical}
      Qualitative behavior of the components of the full and diagonal spectral
      functions $G^{ij}_\rho$ and ${\cal G}^{ii}_\rho$  for $\Gamma_j \ll M_j$
      and $M_j\Gamma_j \ll |\Delta M_{ij}^2|$. For illustration we choose
      $M_2/M_1=2$, $|g_1|/M_1=0.5$ and $|g_2|/M_2=0.8$. The diagonal
      components are \textit{down-scaled} by a factor of 50.
   }
\end{figure}
In Fig.\,\ref{GrhoHierarchical} we show the qualitative behavior 
of the components of the diagonal and full spectral functions as
obtained from Eq.\,\eqref{GFrhoSolution}.
As one can infer from the plot, the off-diagonal components  
$G^{ij}_\rho$ do peak at the same values of the momentum argument
as the diagonal ones. Furthermore, the peaks of $G^{ii}_\rho$ 
and ${\cal G}^{ii}_\rho$  are almost indistinguishable. Therefore 
we can use the Boltzmann equations for the diagonal propagators 
$\hat {\cal G}$ to calculate the asymmetry.

To conclude this section, let us estimate the range of applicability 
of \eqref{EpsilonHierarchical}. It was obtained by approximating the
denominator of \eqref{EpsilonDef} by unity. Since 
$\hat {\cal G}_A^{\dagger}=\hat {\cal G}_R$ and 
$\hat \Pi^{\dagger}_A=\hat\Pi_R$ we obtain
\begin{align}\label{ProdOfDet}
   |g_i|^2 & \left( \Delta_b^i + \Delta_{\bar b}^i \right) \\
           & = \sum_{j\not= i} \bigl[ 2|g_j|^2 | {\cal G}_R^{jj} \Pi_R^{ji} |^2
             - 4\Re\bigl(g_ig_j^*\bigr)\,
             \Re\bigl({\cal G}_R^{jj} \Pi_R^{ji}\bigr) \bigr] \,. \nonumber
\end{align}
Evaluated on the mass-shell of the $i$'th quasiparticle, the retarded
propagator,
\begin{align}
   \label{GRjj}
   {\cal G}^{jj}_R = - \frac{ 1 }{ p^2 - M_j^2 - \Pi_R^{jj} }\,,
\end{align}
is of the order of the inverse splitting of the squared masses, 
$1/\Delta M^2$. Thus the contribution $| {\cal G}_R^{jj} \Pi_R^{ji} |^2$
 is of order of $(\Pi/\Delta M^2)^2$, 
whereas the contribution involving $\Re\bigl({\cal G}_R^{jj} \Pi_R^{ji}\bigr)$ 
is of order of $(\Pi/\Delta M^2)$.

In the hierarchical limit these ratios are much smaller than unity and the
denominator of \eqref{EpsilonDef} can well be approximated by $1$. On the
contrary, in the quasidegenerate case the corrections can become large and
\eqref{EpsilonHierarchical} is no longer applicable.

\subsection{Resonant case}

If the spectrum of heavy (toy-)neutrinos is quasidegenerate,
$|M_1 - M_2|/M_1 \ll 1$, the \CP-violation parameter~\eqref{epsilonclassic}
predicts a resonant enhancement of the generated asymmetry, known
as resonant leptogenesis~\cite{Pilaftsis:2003gt}. Because of  the 
enhancement, this scenario allows  to circumvent the lower bound
on the lightest right-handed neutrino mass $M_1 \gtrsim 10^{9}\mbox{GeV}$ 
typical for thermal leptogenesis~\cite{Davidson:2002qv}.
Therefore, resonant leptogenesis is discussed as 
a possibility to evade constraints from the production of gravitinos 
due to their impact on big bang nucleosynthesis, associated with
the necessity of reheating temperatures well above $M_1$ in the hierarchical
case \cite{Giudice:2008gu}.
Furthermore, it has even been argued that the resonant enhancement could 
lower the scale of right-handed neutrinos to the $\mbox{TeV}$ range, with 
possible implications for collider experiments 
\cite{Blanchet:2008zg,Blanchet:2009bu,Dar:2005hm,Frere:2008ct}.
Therefore, it is important to check whether the conventional Boltzmann treatment
which uses  the canonical \CP-violation parameter~\eqref{epsilonclassic}
agrees with the nonequilibrium field theory description in the quasidegenerate
limit.

Let us start by noting that the result \eqref{EpsilonHierarchical} for the
\CP-violating parameter in the strongly hierarchical limit formally agrees with
the canonical result \eqref{epsilonclassic} even in the quasidegenerate case. 
However, its derivation did involve approximations that break down in the
resonant case. We will now revisit this derivation, starting from
Eq.\,\eqref{EpsilonDef}, without the above approximations. As explained,
the denominator of \eqref{EpsilonDef} can significantly deviate from unity in
the resonant case, which we take into account here. Using Eq.\,\eqref{ProdOfDet} we find 
\begin{equation}
   \label{EpsilonResonantPrel}
   \epsilon_i = \frac{-2\,\Im\bigl(g_j/g_i\bigr)\,
   \Im\bigl({\cal G}_R^{jj} \Pi_R^{ji}\bigr) }
   { 1 + \frac{|g_j|^2}{|g_i|^2} | {\cal G}_R^{jj} \Pi_R^{ji} |^2
   - 2\Re\bigl(g_j/g_i\bigr)\, \Re\bigl({\cal G}_R^{jj} \Pi_R^{ji}\bigr) } \;.
\end{equation}
Note again that we implicitly assume summation over $j\not= i$ in the enumerator
and the denominator, respectively.

From Eqs.~(\ref{ProdOfDet}) and \eqref{PiRADecomposition} we see that the
denominator of \eqref{EpsilonResonantPrel}  involves the self-energy
$\Pi_h^{ij}$, which is logarithmically UV-divergent. It can be renormalized by
including a mass counterterm. As shown in Appendix~\ref{Ren}, this amounts to
the replacements $\hat {\cal G} \rightarrow \hat {\cal G}^{ren}$, and
\begin{align}
   \label{PiRen}
   \hspace*{-1mm}
   \hat \Pi_{R(A)}(X,p) \rightarrow \hat \Pi^{ren}_{R(A)}(X,p)
   \equiv \hat \Pi_{R(A)}(X,p) - \delta\hat Z p^2  + \delta \hat M^2.
\end{align}
For our purposes, it is convenient to use an on-shell renormalization scheme,
for which
\begin{eqnarray}\label{renPrescription}
	\Pi^{ren, ii}_{h, vac}(p^2=M_i^2) & = & 0  \quad (i=1,2)\;,\nonumber\\
	\Pi^{ren, ij}_{h, vac}(p^2=M_i^2) & = & \Pi^{ren, ij}_{h,vac}(p^2=M_j^2)  =  0 \ (i\not= j) \;, \nonumber\\
	\frac{d}{dp^2} \Pi^{ren, ii}_{h, vac}(p^2=M_i^2) & = & 0  \quad (i=1,2)\;, 
\end{eqnarray}
where $\Pi^{ren}_{h, vac}=\Re\Pi^{ren}_{R,vac}=\Re\Pi^{ren}_{A,vac}$ denotes the
dispersive part of the renormalized self-energy in vacuum.

Note that, in vacuum, the self-energy
is time-independent and depends only on $p^2$ due to Lorentz invariance (there
is no medium which singles out a preferred frame).
Note also that the six independent counterterms described by the
symmetric two-by-two matrices $\delta\hat Z$ and $\delta\hat M^2$ are
fully determined by the above renormalization conditions.
For the renormalized self-energy in medium, we find (see
Appendixes~\ref{cpclassic} and~\ref{OneLoopInt})
\begin{eqnarray}
	\Pi^{ren, ii}_h(X,p) & = & \Pi_h^{med, ii}(X,p) \nonumber\\
                             &   & {} + \frac{|g_i|^2}{16\pi^2} \left(
\ln\frac{|p^2|}{M_i^2} - \frac{p^2-M_i^2}{M_i^2}\right) , \; \\
	\Pi^{ren, ij}_h(X,p) & = & \Pi_h^{med, ij}(X,p) \nonumber\\
                             &   & {} +
\frac{\Re(g_ig_j^*)}{16\pi^2}\Bigg[\frac{p^2-M_j^2}{M_i^2-M_j^2}\ln\frac{|p^2|}{
M_i^2} \nonumber\\
                             &   & {} +
\frac{p^2-M_i^2}{M_j^2-M_i^2}\ln\frac{|p^2|}{M_j^2} \Bigg] \ (i\not= j)\;,
\nonumber
\end{eqnarray}
where, in the medium rest-frame (for $i,j=1,2$),
\begin{eqnarray}\label{Lh}
   \Pi_h^{med, ij}(X,p)  & =      & -\frac{\Re(g_ig_j^*)}{16\pi^2|\vec{p}|}
\int_0^\infty dE \; [f_b(E) + f_{\bar b}(E) ] \nonumber\\
                       &        & \times \ln \left| \frac{ (2E+|\vec{p}|)^2 -
p_0^2 }{ (2E-|\vec{p}|)^2 - p_0^2 } \right| \\
                       & \equiv & -\frac{\Re(g_ig_j^*)}{8\pi} L_h^{med}(X,p) \;.
\nonumber
\end{eqnarray}
It is also straightforward to generalize the decomposition~\eqref{PiRADecomposition}
to the renormalized case. While the imaginary part $L_\rho(X,p)$
is not affected, we define (for $i,j=1,2$)

\begin{equation}
   \Pi^{ren, ij}_h(X,p) \equiv {\textstyle \frac{-1}{16\pi}} (g_ig_j^*+g_i^*g_j)
L_h^{ren, ij}(X,p) \;.
\end{equation}

Since we will only use the renormalized quantities from now on, we omit the
superscript `ren' for brevity. Then, using also \eqref{GRADecomposition}, we find
\begin{align}\label{EpsilonResDec}
   \epsilon_i & =  - \frac{|g_j|^2}{16\pi}
                \Im\biggl(\frac{g_i g_j^*}{g_i^* g_j}\biggr) \times \\
              & \frac{ \mathcal{G}_h^{jj}L_\rho + \mathcal{G}_\rho^{jj}L_h^{ij}}
                { 1 + 4 M_j\Gamma_j \frac{ [\Re(g_ig_j^*)]^2 }
                { |g_i|^2 |g_j|^2 } \bigl[ M_j\Gamma_j
                |{\cal G}_R^{jj} L_R^{ji}|^2 + \Re({\cal G}_R^{jj} L_R^{ji})
                \bigr] } \;, \nonumber
\end{align}
where $M_j\Gamma_j\equiv |g_j|^2/(16\pi)$ and
$L_R^{ij} \equiv L_h^{ij} + \frac{i}{2} L_\rho$. Note that ${\cal G}_R^{jj}$,
which is given by \eqref{GRjj}, as well as its real and imaginary components,
${\cal G}_h^{jj}$ and ${\cal G}_\rho^{jj}$, also contain the renormalized
self-energy.

The propagators and the components of the self-energy in \eqref{EpsilonResDec}
are functions of the center coordinate $X$ and the four-momentum $p$. In
general,
the components of $p$ are not related by the on-shell condition, so that
\eqref{EpsilonResDec} describes \CP-violating effects in the decays of 
\textit{on- and off-shell} toy-Majoranas. However, if $|M_i^2-M_j^2|\gtrsim 
M_i\Gamma_i,M_j\Gamma_j$ then the largest contribution to $\epsilon_i$ comes 
from  on-shell momenta.  The in-medium dispersion relation $ E_i^{med}$
is determined by the equation\footnote{The contributions of  
off-shell momenta can be partially taken into account by evaluating the 
\CP-violating parameter at a complex value of $p$
determined by  
$p^2 - M_i^2 - \Pi_R^{ii}(X,p) = 0$ \cite{Anisimov:2005hr}. However, note
that this approach  (just as the original result \eqref{EpsilonResDec}, which is
valid also off-shell) requires the use of an off-shell generalization of the
Boltzmann equations.}

\begin{equation}
   p^2 - M_i^2 - \Re\Pi_R^{ii}(X,p) |_{p^0= E_i^{med}} = 0\;.
\end{equation}
An effective, momentum-dependent mass can be defined by the relation
$E_i^{med} \equiv [(M_i^{med})^2 + |\vec{p}|^2]^\frac12$, which implies
\begin{equation}
   \label{MmedDefinition} 
   M_i^{med} = \bigl( M_i^2 + \Re\Pi_R^{ii}(X,p) |_{p^0= E_i^{med}}
\bigr)^\frac12 \;.
\end{equation}
By means of Eq.\,\eqref{MmedDefinition} 
we  take the mass-shift of the in-medium masses,
$\delta M_i\equiv M_i^{med} - M_i$, into account. In a thermal medium, they
correspond to thermal masses. Even though the mass shifts $\delta M_i$ are
small compared to $M_i$, they can be important for the mass difference
in the quasidegenerate case. 

Finally, we evaluate Eq.\,\eqref{EpsilonResDec} for $p^0= E_i^{med}$, using
Eqs.\,\eqref{GRjj} and \eqref{GRADecomposition}. We obtain the following result
for the \CP-violating parameter:
\begin{align}
   \label{EpsilonResonant}
   \epsilon_i & =  \frac{|g_j|^2}{16\pi} \Im\biggl(\frac{g_i g_j^*}{g_i^*
g_j}\biggr)  \\
              & \times \frac{ \Delta_{ij} - \delta_{ij}  }{ c_{CP}^2 
[\Delta_{ij}- \delta_{ij}]^2 + s_{CP}^2 \left[ \Delta_{ij}^2 + (M_j
\Gamma_j)^2\,L_\rho^2 \right] }\cdot L_\rho \;, \nonumber
\end{align}
where $L_\rho \equiv L_\rho(X,p)|_{p^0= E_i^{med}}$ encodes the medium
correction discussed already in the hierarchical case, obtained
by replacing $M_i\rightarrow M_i^{med}$ in \eqref{LRhoMed}. The denominator
consists of two contributions, weighted by the parameters
\begin{eqnarray}
    c_{CP}^2 & \equiv & [ \Re(g_ig_j^*) / ( |g_i||g_j| ) ]^2  =
\cos^2(\delta_{CP}) \;, \\
    s_{CP}^2 & \equiv & [ \Im(g_ig_j^*) / ( |g_i||g_j| ) ]^2  =
\sin^2(\delta_{CP}) \;, \nonumber
\end{eqnarray}
related to the \CP-violating phase $\delta_{CP}\equiv \arg(g_1^*g_2)$.
Furthermore, the degeneracy parameter,
\begin{equation}
   \Delta_{ij} \equiv (M_i^{med})^2 - (M_j^{med})^2 + \delta_{ij}'  \;,
\end{equation}
is given by the difference of the \emph{in-medium} (``thermal'') masses of the
heavy toy-neutrinos, plus logarithmic corrections (here we display the
superscript `$ren$' again for clarity):
\begin{eqnarray}
   \delta_{ij}  & \equiv & 2 M_j\Gamma_j L_h^{ren, ij}(X,p)|_{p^0=E_i^{med}} \;,\\
   \delta_{ij}' & \equiv & 2 M_j\Gamma_j \Big( L_h^{ren,
jj}(X,p)|_{p^0=E_i^{med}} \nonumber\\
                &        & \qquad\qquad - L_h^{ren, jj}(X,p)|_{p^0=E_j^{med}}\Big) \;. \nonumber
\end{eqnarray}

As we have already mentioned, the resonance effects modify not only the
\CP-violating parameters, but also the total decay widths of the heavy particles; see
Eq.\,\eqref{GammaInMedium}. Using the functions and definitions 
introduced above we can write the in-medium decay widths in the form
\begin{align}
&\Gamma_i^{med}=\Gamma_i\times\\
&\frac{c_{CP}^2  [\Delta_{ij}- \delta_{ij}]^2 + s_{CP}^2 \left[ \Delta_{ij}^2 +
(M_j \Gamma_j)^2\,L_\rho^2 \right] }{
[\Delta_{ij}-2c^2_{CP}\delta_{ij}]^2+[s_{CP}^2(M_j\Gamma_jL_\rho)
+c^2_{CP}\delta^2_{ij}/(M_j\Gamma_j L_\rho)]^2
}.\nonumber
\end{align}

In the limit $|M_i^2-M_j^2| \gg M_j\Gamma_j$, the \CP-violating
parameter~\eqref{EpsilonResonant} converges toward the result
\eqref{EpsilonHierarchical} for the hierarchical case.
Our result for the resonant case, Eq.\,\eqref{EpsilonResonant}, 
describes the leading corrections when the above ratio becomes sizable.
However, one should  keep in mind that, in the ``maximal resonant'' 
case, where the decay width is comparable to the mass difference, 
the Boltzmann picture breaks down. The following simple argument 
supports this statement: To a good 
approximation the spectral functions ${\cal G}_\rho^{ii}$ of the heavy 
fields have a Breit-Wigner shape. To evaluate the gain and loss 
terms \eqref{SigmasWT} we have to integrate over the frequency
$k^0$ of the heavy particles' propagator. If the distance between
the peaks of the spectral function is considerably larger than the
decay widths, the integration reduces to two independent integrations
in the vicinities of the corresponding mass shells. Thus, we can
identify two independent quasiparticle excitations with the 
corresponding distribution functions $f_{\psi_1}$ and  $f_{\psi_2}$
and the \CP-violating parameters $\epsilon_1$ and $\epsilon_2$ 
which contribute to the generation of the asymmetry. On the 
other hand, if the decay width is comparable to the difference of 
the masses, then the peaks of the spectral functions overlap and 
neither the distribution functions, nor the \CP-violating 
parameters, are well-defined. 
\begin{figure}[h!]
   \includegraphics[width=0.90\columnwidth]{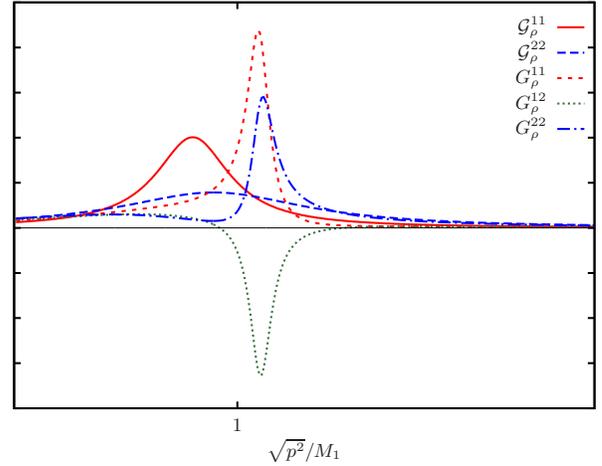} 
   \caption{\label{ResonantProp}
      Qualitative behavior of the diagonal 
      and off-diagonal components of full and diagonal spectral functions
      $G_\rho^{ij}$ and ${\cal G}_\rho^{ii}$ in the maximal resonant limit
      $M_j^2 \gg |\Delta M_{ij}^2| \lesssim M_j\Gamma_j$. For illustration
      we choose $M_2/M_1=1.02$, $|g_1|/M_1=0.5$ and $|g_2|/M_2=0.8$. 
      The diagonal components are \textit{not} scaled.
   }
\end{figure}
In Fig.\,\ref{ResonantProp} we present the qualitative 
behavior of the diagonal and off-diagonal components of the 
full and diagonal  spectral functions  $G_\rho^{ij}$ and
${\cal G}_\rho^{ii}$.
As one can infer from the plot, the off-diagonal components 
of the full spectral function no longer have two pronounced 
peaks. Thus, it is not possible to define two \CP-violating
parameters. The shape of the diagonal components deviates 
from the Breit-Wigner one, and the positions of the peaks
of $G_\rho^{ii}$ are shifted as compared to those for the 
diagonal spectral function ${\cal G}_\rho^{ii}$. In other 
words, using the Boltzmann equation for the diagonal 
propagators $\hat{\cal G}$ to calculate the generated asymmetry
is no longer a good approximation because of the off-shell effects.
Moreover, since the microscopic time scales
$(t_{mic} \sim \Delta M^{-1})$ and the macroscopic time scales $(t_{mac} \sim 
\Gamma^{-1})$ can be of the same order of magnitude in the ``maximal resonant'' 
regime, the memory effects can play an important role. Thus we conclude that in
this case, one should use the
quantum kinetic equations discussed in Appendix~\ref{real}, which take
off-shell effects into account, or even the full nonlocal quantum evolution
equations discussed in Appendix~\ref{FullKB}, which also capture memory and
correlation effects.

With these considerations in mind, let us now analyze the
Kadanoff--Baym/Schwinger--Keldysh result \eqref{EpsilonResonant} for the
\CP-violating parameter in the resonant (but not maximal resonant) regime, for
some particular cases of interest. For definiteness we concentrate here on
$\epsilon_1$, i.e. $i=1,j=2$ assuming that $M_1<M_2$. In terms of the ratio
\begin{equation}
   R \equiv \frac{M_j^2 - M_i^2}{M_j\Gamma_j}\,,
\end{equation}
the condition of validity for the Boltzmann-treatment can be expressed as 
$R \gg 1$.
The hierarchical limit corresponds to $R \rightarrow \infty$.
In the following we study the leading corrections in $1/R$.

In order to separate medium and resonance corrections, we first consider
the vacuum limit $f_{b(\bar b)} \rightarrow 0$. The renormalization
prescription Eqs.\,\eqref{renPrescription} ensures that
$M_i^{med} \rightarrow M_i^{vac} = M_i$, i.e. $M_1$ and $M_2$ are the on-shell
masses in vacuum. Furthermore, adopting the scheme \eqref{renPrescription}
also implies that $\delta_{ij} \rightarrow \delta_{ij}^{vac} = 0$ and
\begin{equation}
   \delta_{ij}' \rightarrow (\delta_{ij}')^{vac} =
-\frac{|g_j|^2}{16\pi^2}\left[ \ln\left(\frac{M_i^2}{M_j^2}\right) -
\frac{M_i^2-M_j^2}{M_j^2}\right]\;.
\end{equation}

Therefore, in the vacuum limit the degeneracy parameter is given by
\begin{equation}
   \Delta_{ij} \rightarrow \Delta^{vac}_{ij} = M_i^2 - M_j^2 +
(\delta_{ij}')^{vac}\;.
\end{equation}

For $|g_j|/M_j\ll 1$, the loop correction can be safely neglected.
Thus, in the vacuum limit the \CP-violating parameter is approximately given by
\begin{align}\label{EpsilonResonantVac}
   \epsilon_i^{vac} & \approx - \frac{|g_j|^2}{16\pi} \Im\biggl(\frac{g_i
g_j^*}{g_i^* g_j}\biggr) \frac{ M_j^2 - M_i^2  }{ (M_j^2 - M_i^2)^2 + s_{CP}^2
M_j^2 \Gamma_j^2 } \nonumber\\
   & = \sin(2\delta_{CP}) \frac{R}{ R^2 + \sin^2(\delta_{CP}) }\,.
\end{align}
Compared to the conventional result \eqref{epsilonclassic}  
the width in the denominator is effectively changed according to 
$\Gamma_j \rightarrow |s_{CP}|\Gamma_j$. Since $|s_{CP}| \leq 1$, this 
means that for fixed values of $R$ and $\delta_{CP}$ the result above 
is always larger compared to the conventional one. This property may be 
attributed to the fact that the Kadanoff--Baym formalism takes a 
resummation of resonant contributions into account. Note that, if 
taken at face value, the expression \eqref{EpsilonResonantVac} for
$\epsilon_i^{vac}$ formally has a peak at $R_* \sim |s_{CP}|$, with 
maximum $|\epsilon_i^{*}|=|c_{CP}|$ and width $\sim |s_{CP}|\Gamma_j$.
The width reduces to zero in the limit of vanishing \CP-violation,
$\delta_{CP}\rightarrow 0$, although the peak value remains finite. 
On the contrary, for the conventional result \eqref{epsilonclassic} 
one has $R_* \sim 1$, $|\epsilon_i^{*}|=|s_{CP}c_{CP}|$ and
a width $\sim \Gamma_j$. However, we stress that the result should only 
be trusted in the regime where $R \gg 1$, and can be modified significantly 
by medium effects.

There are two sources of medium corrections in the resonant case. The first is
the contribution from the spectral (imaginary) part of the self-energy loop,
given by $L_\rho$ in Eq.\,\eqref{EpsilonResonant}. This contribution is known 
already from the hierarchical case; see Eq.\,\eqref{EpsilonHierarchical} as 
well as Fig.\,\ref{thermalEpsilonAveraged}. The second contribution stemming 
from the real part of the self-energy loop, given by $L_h^{med}$, enters via
the thermal masses and also via the ``logarithmic'' corrections.

For illustration, we insert a Bose-Einstein distribution for $f_b$ and $f_{\bar
b}$ (with $\mu=0$) in the expression for $\Pi_h^{med}$ in Eq.\,\eqref{Lh}. 
In the limit $T \lesssim M_i$, we obtain
\begin{equation}
   (M_i^{med})^2  \approx  M_i^2 + {\textstyle \frac{1}{6}\frac{|g_i|^2}{M_i^2}
} \,T^2 \,,
\end{equation}
\begin{eqnarray}\label{deltaThermal}
   \delta_{ij}  & \approx & - {\textstyle \frac{|g_j|^2}{16\pi^2} } \ln\left(
\frac{ (M_i^{med})^2 }{ M_i^2 }  \right) - {\textstyle
\frac{1}{6}\frac{|g_j|^2}{M_i^2} } \,T^2 \,, \nonumber\\
   \delta_{ij}' & \approx & - {\textstyle \frac{|g_j|^2}{16\pi^2} } \left[
\ln\left( \frac{ (M_i^{med})^2 }{ (M_j^{med})^2 }  \right) - \frac{
(M_i^{med})^2 - (M_j^{med})^2 }{ M_j^2 } \right] \nonumber\\
                &         & {} - {\textstyle \frac{1}{6}}|g_j|^2 \left(
(M_i^{med})^{-2} - (M_j^{med})^{-2}\right) T^2\,. 
\end{eqnarray}
The term in square brackets is of second order in the mass-squared difference
and can be neglected. Assuming in addition $|g_j|^2\ll M_j^2$, the
degeneracy parameter is approximately given by
\begin{equation}\label{DeltaThermal}
   \Delta_{ij} \approx M_i^2 - M_j^2 + \frac{T^2}{6} \left(
\frac{|g_i|^2}{M_i^2} - \frac{|g_j|^2}{M_j^2} \right) \,.
\end{equation}
In the quasi degenerate limit it is plausible to assume that also 
the couplings are quasi degenerate, $|g_1|^2\simeq |g_2|^2$. In 
this case, the thermal corrections stemming from the dispersive part of the
self-energy cancel approximately. In general, the thermal corrections 
could also be larger than the mass-splitting of the vacuum masses. 
If the vacuum masses are quasi degenerate, this would destroy the 
resonance condition. However, also the opposite case
is possible, namely $\Delta_{ij}$ could become tiny due to a cancellation of
the vacuum masses and the thermal corrections for a certain temperature.
Then the resonant enhancement would occur only close to this particular
temperature. We do not pursue these possibilities further here.
Assuming, for simplicity, that the thermal 
contributions in Eqs.\,\eqref{deltaThermal} and \eqref{DeltaThermal} are 
subdominant, we can summarize the results for the medium- and resonance 
corrections to the \CP-violating parameter as follows:
\begin{align}\label{EpsilonComparison}
   \epsilon_i \approx \sin(2\delta_{CP}) \left\{
   \begin{array}{cl}
      \displaystyle \frac{R}{ R^2 + 1 } & \ \mbox{conventional} \\[3ex]
      \displaystyle \frac{R\cdot L_\rho}{ R^2 + L_\rho^2 } & \ \mbox{med. corr.}
\\[3ex]
      \displaystyle \frac{R\cdot L_\rho}{ R^2 + s_{CP}^2 L_\rho^2 } & \
\mbox{med.+res. corr.}
   \end{array}
   \right.
\end{align}
The first two expressions follow directly from Eq.\,\eqref{epsilonclassic} 
and Eq.\,\eqref{EpsilonHierarchical}, respectively. The third one approximates 
the resonant result, Eq.\,\eqref{EpsilonResonant}, for $R \gtrsim 5$.

\begin{figure}[h!]
   \includegraphics[width=\columnwidth]{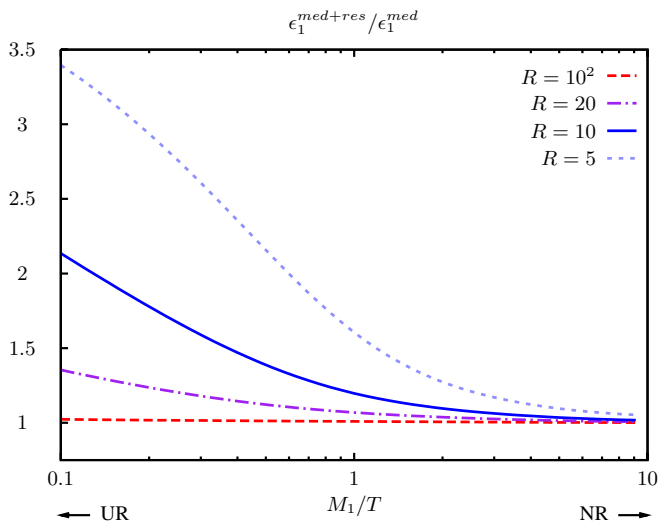}
   \caption{\label{epsilonVsInverseTemperature}%
      Corrections to the effective self-energy \CP-violating parameter  
      $\epsilon_1$ in the resonant case obtained from the Kada\-noff--Baym 
      formalism (for $|g_1| = |g_2| = 0.01M_1$, $\delta_{CP}=\pi/8$ and
       $|\vec{p}|=T$; the mass $M_2$ is determined by the value of $R$).
   }
\end{figure}

In Fig.\,\ref{epsilonVsInverseTemperature} we show the ratio of the expressions
for the CP-violating parameter, Eq.\,\eqref{EpsilonResonant}, 
which includes medium and resonance corrections, and  
Eq.\,\eqref{EpsilonHierarchical}, which includes only medium corrections, 
for several values of the degeneracy parameter $R$. 
Large values of $R$ correspond to the hierarchical case and 
there is no difference between the two approximations. On the other hand, 
for small values of $R$ the corrections are substantial. For instance
for $R= 10$ the  expression  without resonance corrections underestimates the
\CP-violating parameter by a factor of two at high temperatures.

Since typically most of the asymmetry is generated at $T\sim M_1$, it is 
instructive to look at the \CP-violating parameter as a function of the 
degeneracy parameter $R$ at $T=M_1$. 
\begin{figure}[ht!]
   \includegraphics[width=\columnwidth]{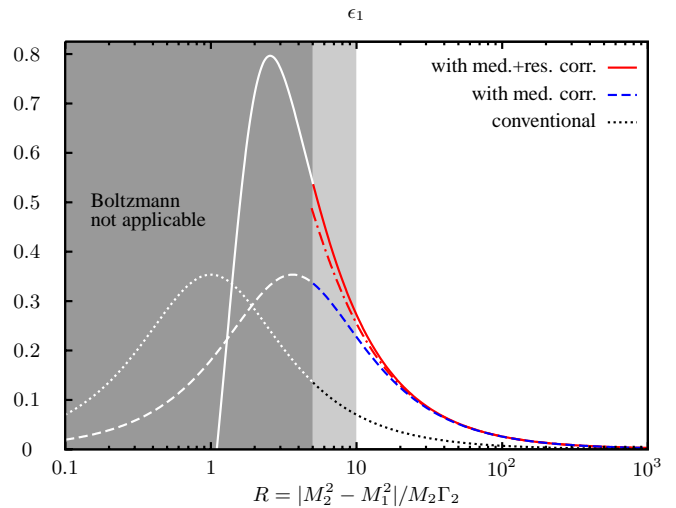}
   \caption{\label{figures/epsilonVsR}%
                Dependence of the vacuum (dotted line), hierarchical (dashed line) and
      resonant (solid line) approximations for the \CP-violating parameter on
      the degeneracy parameter $R$ calculated at $T=M_1$ (for the same
      parameter values as in Fig.\,\ref{epsilonVsInverseTemperature}). The
      dot-dashed line shows the approximate expression for the resonant case
      from the last line of Eq.\,\eqref{EpsilonComparison}. The
      Boltzmann-approximation requires $R \gg 1$ and therefore the results 
      for the \CP-violating parameter are not applicable in the gray shaded 
      region. We show them only for comparison with the conventional result.
   }
\end{figure}

In Fig.\,\ref{figures/epsilonVsR} we present the $R$-dependence of
the conventional vacuum approximation for the \CP-violating parameter, 
Eq.\,\eqref{epsilonclassic}, the hierarchical approximation  in medium,
Eq.\,\eqref{EpsilonHierarchical}, and the resonant expression,
Eq.\,\eqref{EpsilonResonant}, respectively. Very large values of $R$ correspond
to $M_2\gg M_1$. In this case the resonance effects are suppressed
and all three expressions go to zero $\propto R^{-1}$. For 
smaller values of $R$ we observe a significant deviation of the \CP-violating 
parameter from its vacuum value, which is due to medium 
effects. Finally for even smaller values of $R$ the resonant effects
become important and we observe a deviation of the \CP-violating 
parameter from its value calculated in the hierarchical approximation.

\begin{figure}
\includegraphics[width=\columnwidth]{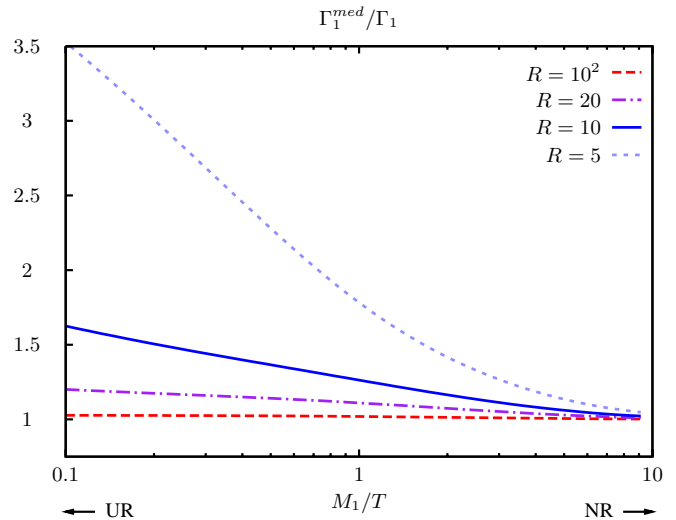}
\caption{\label{InMedWidthVsTPlot} Ratio of the in-medium decay width of the 
lightest toy-Majorana to its tree-level value as a function of the 
dimensionless inverse temperature calculated for various values of the 
degeneracy parameter $R$. 
}
\end{figure}

\begin{figure}
\includegraphics[width=\columnwidth]{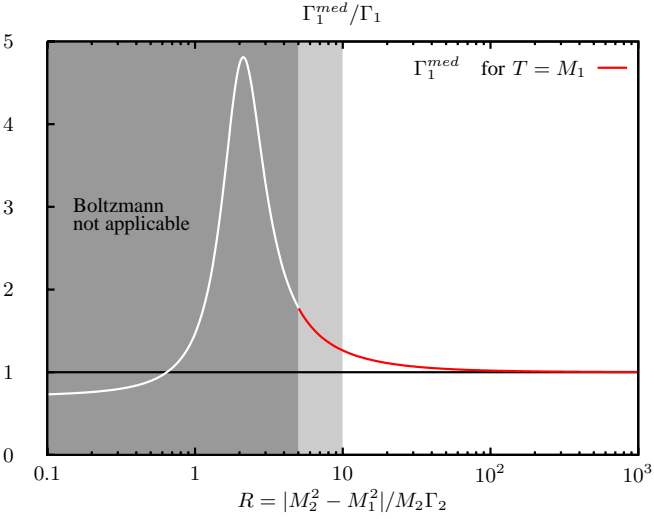}
\caption{\label{InMedWidthVsRPlot} Ratio of the in-medium decay width of the 
lightest toy-Majorana to its tree-level value as a function of the 
degeneracy parameter $R$ calculated at $T=M_1$. 
}
\end{figure}

The effective decay widths $\Gamma_i^{med}$ are also enhanced by the medium 
and resonance effects. The enhancement increases
with the temperature; see Fig.\,\ref{InMedWidthVsTPlot}.
Furthermore, it strongly depends on the values of the
degeneracy parameter $R$, as can be seen in Fig.\,\ref{InMedWidthVsRPlot}.
Even for reasonable values of the degeneracy parameter the effective decay 
width can be twice as large as in vacuum. This leads to a faster decay of the 
toy-Majoranas. Also the inverse decay processes are more efficient in washing 
out the asymmetry. In other words, the increase in the generated asymmetry 
due to the enhancement of the \CP-violating parameter can be partially 
compensated by the increase of the in-medium decay widths. 
Let us also note that the observed enhancement of the total in-medium decay 
widths could be very important in the scenarios with very small values of the 
degeneracy parameter \cite{Boyarsky:2009ix,Shaposhnikov:2008pf}.

% =============================================================================
\section{\label{Numerics}Numerical results}
% =============================================================================

For the moment we consider numerical solutions only for the strongly
hierarchical case, as the resonant case is more involved and one cannot obtain
consistent Boltzmann equations for the maximal resonant case. To obtain the
Boltzmann equations for $\f{b}{}$ and  $\f{\bbar}{}$ we integrate
\eqns\eqref{BlzmnClassic} and the corresponding equation for $\bbar$, together
with  Eq.\,\eqref{SigmasWT}, over $p_0$. The Boltzmann equations for
$f_{\psi_i}$ are obtained from \eqn\eqref{GFBlzmn}. As one can infer from
\fig\ref{GrhoHierarchical}, in the hierarchical case the off-diagonal components
of the full propagators $\hat G$ are subdominant and the diagonal components of
$\hat G$
are almost identical to those of $\hat {\cal G}$. Therefore, we can neglect the
off-diagonal components in the kinetic equations \eqref{GFBlzmn} for the
full propagators  $\hat G_{\gtrless}$  and approximate them by the kinetic
equations \eqref{GFrhodiagBlzmn} for the corresponding diagonal propagators
${\hat {\cal G}}_\gtrless$. The Boltzmann equations are then obtained after
using the Kadanoff--Baym ansatz \eqref{Giigtrless} and the quasiparticle
approximation \eqref{GrhoQP}.

We solve the coupled system of Boltzmann equations in the spatially homogeneous
and isotropic case in (spatially flat and radiation dominated)
Friedman--Robertson--Walker space-time. They take the form
\begin{align}
   \lorig [f](\momabs)\equiv&p^\alpha {\cal D}_\alpha f(\momabs)\nonumber\\
   = & p^0\left(\frac{\partial }{\partial t}-\momabs\hubblerate
   \frac{\partial }{\partial \momabs}\right)f(\momabs)\,,
\end{align}
where $\hubblerate\equiv {\dot{a}}/{a}$ is the Hubble parameter. As usual, the
integrations over the time components of each of the invariant four-volume
elements in the collision terms can be performed trivially after the
quasiparticle approximations for the spectral functions have been inserted.

\begin{widetext}
   The resulting system of Boltzmann equations takes the same form as the one
   presented in \cite{Garny:2009rv} for the vertex contributions. As in the
   vertex case the structure differs from the usual one obtained in the
   conventional bottom-up approach. In particular, we do not need to include
   the RIS part of the collision terms for the processes
   $bb\leftrightarrow \bbar \bbar $ because our collision terms for the
   processes $bb\leftrightarrow\psi_1$ and $\bbar \bbar \leftrightarrow\psi_1$
   do not suffer from the  generation of an asymmetry in equilibrium. The form
   of these equations is also necessary to guarantee cancellation of the
   gain- and loss-term contributions in equilibrium when the quantum statistical
   terms are present. This structure can be translated directly from the
   toy-model to established scenarios of leptogenesis and baryogenesis by
   analogy. Therefore, we consider it as important, also for phenomenological
   studies, and repeat it here:
   \begin{subequations}
      \label{QuantumBoltzmannEquations}
      \begin{align}
         \label{boltzmann equation b}
         \lorig [\f{b}{}](\momkabs)&=\carrayorigkb{bb\leftrightarrow
bb}{\momkabs}{\f{b}{}}+
         \carrayorigkb{b\bbar \leftrightarrow b\bbar
}{\momkabs}{\f{b}{},\f{\bbar}{}}+ 
         \carrayorigkb{bb\leftrightarrow
\psi_1}{\momkabs}{\f{b}{},\f{\psi_1}{}}\kend\\
         \label{boltzmann equation bbar}
         \lorig [\f{\bbar}{}](\momkabs)&=\carrayorigkb{\bbar \bbar
\leftrightarrow 
         \bbar \bbar }{\momkabs}{\f{\bbar}{}}+\carrayorigkb{\bbar
b\leftrightarrow 
         \bbar b}{\momkabs}{\f{\bbar}{},\f{b}{}}+\carrayorigkb{\bbar \bbar
\leftrightarrow 
         \psi_1}{\momkabs}{\f{\bbar}{},\f{\psi_1}{}}\kend\\
         \label{boltzmann equation psi}
         \lorig [\f{\psi_1}{}](\momkabs)&=\carrayorigkb{\psi_1\leftrightarrow
{b}{b}}{\momkabs}{\f{\psi_1}{},\f{b}{}}
         +\carrayorigkb{\psi_1\leftrightarrow \bbar \bbar
}{\momkabs}{\f{\psi_1}{},\f{\bbar}{}}\dend
      \end{align}
   \end{subequations}
   Here, $\carrayorigkb{i\leftrightarrow f}{}{}$ denotes the collision term
   for a process $i\rightarrow f$. The collision terms for the $2-2$ scattering
   processes in \eqref{boltzmann equation b} are given by
   \begin{subequations}
      \label{2bodyscattering}
      \begin{align}
         \label{collsion term bb-bb}
         \carrayorigkb{bb\leftrightarrow bb}{\momkabs}{\f{b}{}} =
\textstyle{\frac{1}{2}}\textint & \lorentzd{3}{\momp} \lorentzd{3}{\momq}
\lorentzd{3}{\momr} (2\pi)^4
\delta^{}(p+{\momp}-{\momq}-{\momr}){\textstyle{\frac{1}{2}}}\lambda^2
\nonumber\\
         \times
&\big\{\qstat{\f{b}{{\momkabs}}}\qstat{\f{b}{{\mompabs}}}\f{b}{{\momqabs}}\f{b}{
{\momrabs}}-\f{b}{{\momkabs}}\f{b}{{\mompabs}}\qstat{\f{b}{{\momqabs}}}\qstat{\f
{b}{{\momrabs}}}\big\}\kend\\
         \hspace{-2mm}
         \label{collsion term bbbar-bbbar}\carrayorigkb{b\bbar \leftrightarrow
b\bbar
}{\momkabs}{\f{b}{},\f{\bbar}{}}=\textstyle\textstyle{\frac{1}{2}}\textint &
\lorentzd{3}{\momp} \lorentzd{3}{\momq} \lorentzd{3}{\momr}(2\pi)^4
\delta^{}(p+{\momp}-{\momq}-{\momr})\lambda^2\nonumber\\
         \times
&\big\{\qstat{\f{b}{{\momkabs}}}\qstat{\f{\bbar}{{\mompabs}}}\f{\bbar}{{\momqabs
}}\f{b}{{\momrabs}}-
\f{b}{{\momkabs}}\f{\bbar}{{\mompabs}}\qstat{\f{\bbar}{{\momqabs}}}\qstat{\f{b}{
{\momrabs}}}\big\}\dend
      \end{align}
   \end{subequations}
   Replacing $\f{b}{}$ with $\f{\bbar}{}$ in \eqns\eqref{2bodyscattering} one
   obtains the analogous terms in the equation for $\bbar$. The collision terms
   in \eqn\eqref{boltzmann equation psi} are obtained by inserting the diagonal
   components of the self-energy \eqn\eqref{Pi} into
   \eqn\eqref{GFrhodiagBlzmn}:
   \begin{align}
      \label{collision term psi-bb and psi-bbar bbar}
      \carrayorigkb{\psi_1\leftrightarrow bb}{\momkabs}{\f{\psi_1}{},\f{b}{}}+ &
\carrayorigkb{\psi_1\leftrightarrow \bbar \bbar
}{\momkabs}{\f{\psi_1}{},\f{\bbar}{}} \nonumber\\
      \simeq\textstyle{\frac{1}{2}}\textint & \lorentzd{3}{\momp}
\lorentzd{3}{\momq} (2\pi)^4
\delta^{}(p-{\momp}-{\momq}){\textstyle{\frac{1}{2}}}|g_1|^2\nonumber\\ 
      \times \big[& \big\{
\qstat{\f{\psi_1}{{\momkabs}}}\f{b}{{\mompabs}}\f{b}{{\momqabs}}-\f{\psi_1}{{
\momkabs}}\qstat{\f{b}{{\mompabs}}}\qstat{\f{b}{{\momqabs}}}\big\}\nonumber\\
+&\big\{\qstat{\f{\psi_1}{{\momkabs}}}\f{\bbar}{{\mompabs}}\f{\bbar}{{\momqabs}}
-\f{\psi_1}{{\momkabs}}\qstat{\f{\bbar}{{\mompabs}}}\qstat{\f{\bbar}{{\momqabs}}
}\big\}\big]\dend
   \end{align}
   In the framework of the toy model the \CP-violating parameter for the
   self-energy loop contributions $\epsilon_i$ given in
   \eqn\eqref{EpsilonHierarchical}, in the strongly hierarchical limit
   $M_2 \gg M_1$, differs by just a (symmetrization) factor $1/2$ from the
   vertex contributions. It appears explicitly in the collision terms for the
   (inverse) decay of $\psi_1$ into $bb$ or $\bbar  \bbar $:
   \begin{subequations}
      \label{collision_term_bbbarbbarb-psi}
      \begin{align}
         \label{collision term bb-psi}
         \carrayorigkb{bb\leftrightarrow
\psi_1}{\momkabs}{\f{b}{},\f{\psi_1}{}}=\textstyle{\frac{1}{2}}\textint &
\lorentzd{3}{\momp} \lorentzd{3}{\momq} (2\pi)^4 \delta^{}({\momp}-p-{\momq})
|g_1|^2[1+\epsilon_1(\momqabs)]\nonumber\\
         \times &
\big\{\qstat{\f{b}{{\momkabs}}}\qstat{\f{b}{{\mompabs}}}\f{\psi_1}{{\momqabs}}
-\f{b}{{\momkabs}}\f{b}{{\mompabs}}\qstat{\f{\psi_1}{{\momqabs}}}\big\}\kend\\
         \label{collision term bbar bbar-psi}
         \carrayorigkb{\bbar \bbar \leftrightarrow
\psi_1}{\momkabs}{\f{\bbar}{},\f{\psi_1}{}}=\textstyle{\frac{1}{2}}\textint &
\lorentzd{3}{\momp} \lorentzd{3}{\momq} (2\pi)^4
\delta^{}({\momp}-p-{\momq})|g_1|^2[1-\epsilon_1(\momqabs)]\nonumber\\
         \times &
\big\{\qstat{\f{\bbar}{{\momkabs}}}\qstat{\f{\bbar}{{\mompabs}}}\f{\psi_1}{{
\momqabs}}-\f{\bbar}{{\momkabs}}\f{\bbar}{{\mompabs}}\qstat{\f{\psi_1}{{\momqabs
}}}\big\}\dend
      \end{align}
   \end{subequations}
\end{widetext}
%~\newpage~\newpage
The network of Boltzmann equations \eqref{QuantumBoltzmannEquations} should be
understood in a generalized sense. The ``amplitudes'' which appear here differ 
from the usual perturbative matrix elements and do not share their symmetry 
properties.

In order to study the effect of the quantum corrections, we can again compare 
the results obtained by integrating the network of Boltzmann equations with 
quantum-corrected $\epsilon_1(\left|\bvec{p}\right|)$ with the corresponding 
ones in the vacuum limit $\epsilon_1^{vac}$. The computation is started at 
sufficiently high temperatures so that all species, including $\psi_1$ with 
mass $M_1=\Cmasspsi\,\mbox{GeV}$, have relativistic initial distributions. 
In addition, we assume that the interactions are in chemical equilibrium in 
the beginning, i.e.~$\mu_{\psi_1}=2\mu_b=2\mu_{\bbar }$. 
We start with sufficiently negative chemical potentials $\mu_b$ as to avoid
Bose--Einstein condensation of the different species\footnote{As was explained 
in \cite{Garny:2009rv} this necessity arises here, because we consider the
system \eqref{QuantumBoltzmannEquations}-\eqref{collision_term_bbbarbbarb-psi}
as closed so that there are no interactions which can remove the produced
over-densities of $b$'s and $\bbar$'s from the system. Therefore $b$ and
$\bbar$ can in principle undergo Bose--Einstein condensation, which we avoid by
choosing $\mu_b$ and $\mu_{\bbar}$ appropriately. Such interactions will be
present in a phenomenological scenario. Whether the possibility of
Bose--Einstein condensation exists in such scenarios will have to be answered
by solving appropriate kinetic equations.}.

\begin{figure}[t]
   \begin{center}
      \includegraphics[width=\columnwidth]{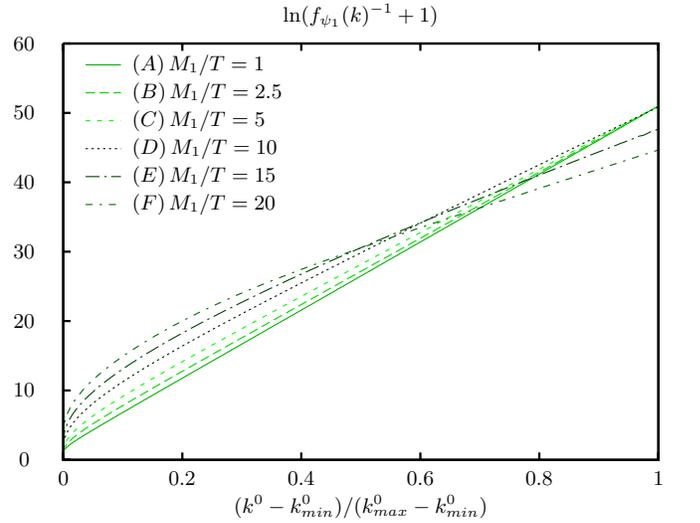}
   \end{center}
   \caption{\label{lnfvsen_02_06_13}
      The distribution function for the massive species $f_{\psi_1}$ can
      deviate significantly from equilibrium for smaller washout factors
      (here $\kappa\simeq\Ckappaa$).
   \protect}
\end{figure}

We choose the coupling $\lambda$ (and $g_i$ via $\epsilon_1$ and $\kappa$) such
that the $2-2$ scattering rates are much larger than those of the decays and
inverse decays. This assures that the light species are kept in kinetic
equilibrium, as  in the standard leptogenesis scenario. As shown in
\cite{Garny:2009rv} there is no need to compute the collision integrals for
$2-2$ scattering explicitly in this case. This means that they can be described
in terms of four parameters $\mu_b$, $T_b$ and $\mu_{\bbar}$, $T_{\bbar}$ which
obey the relation $T_{\bbar}=T_b$. Hence, we studied the evolution of $f_b$ and
$f_{\bar b}$ in terms of only three parameters. In contrast, the full equation
for ${\psi_1}$ was discretized on a grid with $\Cdim$ momentum modes and
solved simultaneously. For this purpose, homogeneity and isotropy was assumed,
so that the angular integration can be performed as in \cite{hohenegger:063502}
in a rather general case or in the appendix of \cite{Garny:2009rv} for the
present special case. As shown in \fig\ref{lnfvsen_02_06_13}, for small washout 
parameters, 
\begin{equation}
   \kappa\equiv \Gamma_1/H(M_1)=|g_1|^2m_{pl}/(4.5\cdot 16\pi\sqrt{g_*}M_1^3)\;,
\end{equation}
the distribution function can deviate significantly from the equilibrium 
form (for which the curves would be straight lines). An equilibrium form 
would be a necessary assumption to obtain rate equations.
The generated ``baryon'' asymmetry is defined as
\begin{equation}
   \eta(M_1/T) = \frac{n_b(M_1/T)-n_{\bbar }(M_1/T)}{s(M_1/T) }\kend
\end{equation}
where $n_{b}$ and $n_{\bbar }$ are the number densities of species $b$ 
and $\bbar $ and $s$ is the standard cosmological entropy 
density~\cite{Kolb:1990vq}. We denote the analogous quantity, corresponding 
to the solution for $\epsilon^{vac}_1$, by $\eta^{vac}(M_1/T)$.
\begin{figure}[t]
   \begin{center}
      \includegraphics[width=\columnwidth]{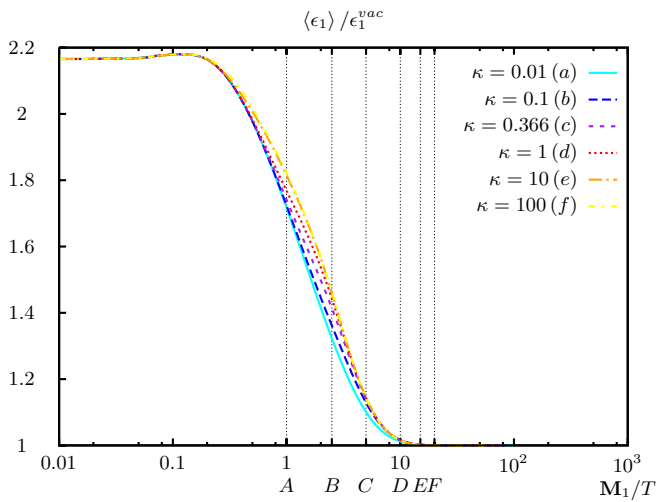}
   \end{center}
   \caption{\label{epsilonepsilonvacvsmt} 
      The ratio $\left<\epsilon_1\right>/\epsilon_1^{vac}$. 
      The curves flatten for small $M_1/T$ because the initial
      conditions involve a finite chemical potential.
   }
\end{figure} 
\Fig\ref{epsilonepsilonvacvsmt} shows the result for the ratio
$\left<\epsilon_1\right>/\epsilon_1^{vac}$  for various values of the washout
parameter. Comparing it to the thermal equilibrium result in
\fig\ref{thermalEpsilonAveraged} one sees a flattening for small $M_1/T$ which
is caused by the finite chemical potential of $b$ and $\bbar $ in the initial
conditions. One would obtain larger corrections if additional interactions for
$b$ and $\bbar $ would be introduced in order to start with smaller chemical
potentials.

\begin{figure}[t]
   \begin{center}
      \includegraphics[width=\columnwidth]{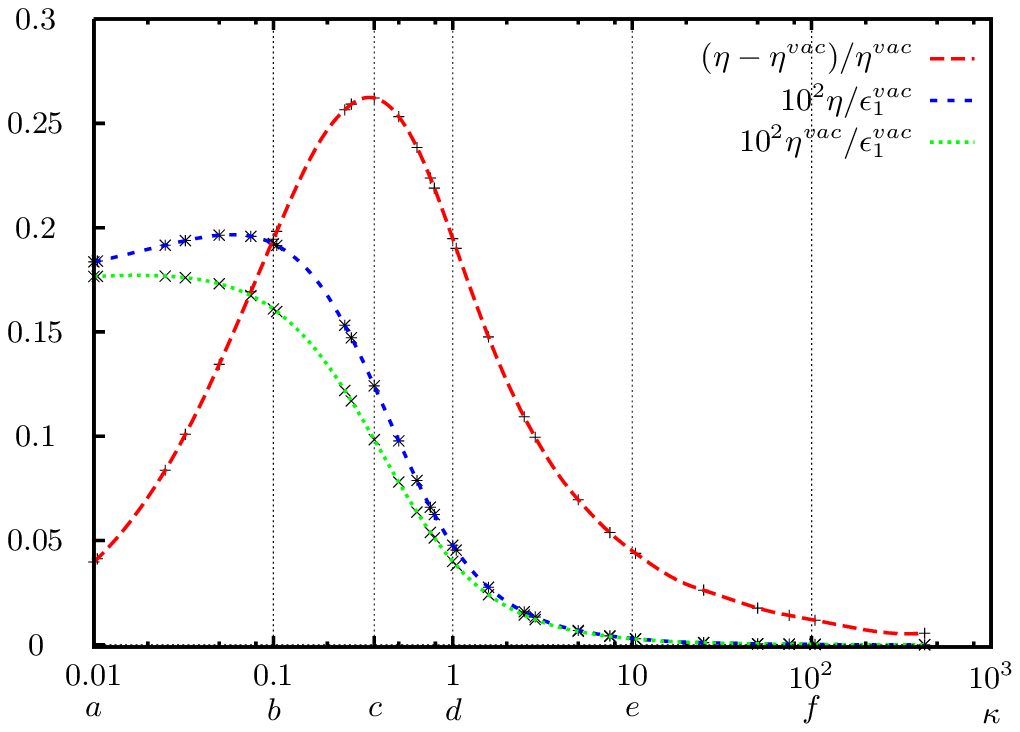}
   \end{center}
      \caption{\label{etavswashout}
      The final asymmetries and the relative quantum correction
      $(\eta-\eta^{vac})/\eta^{vac}$ in dependence of the washout factor
      $\kappa$. The cases $a$, $b$, $c$, $d$, $e$, $f$ correspond to washout
      factors $\Ckappaa$, $\Ckappab$, $\Ckappac$, $\Ckappad$, $\Ckappae$,
      $\Ckappaf$.
   }
\end{figure}

The generated ``baryon'' asymmetry does not depend mo\-no\-to\-no\-usly 
on the washout parameter $\kappa$ when the medium corrections  are taken 
into account. This can be inferred from \fig\,\ref{etavswashout} where the 
dependence of the final asym\-metries $\eta =\eta(M_1/T\rightarrow \infty)$ 
and $\eta^{vac}=\eta^{vac}(M_1/T\rightarrow \infty)$ are presented.
In the present case, where $b$ and $\bbar$ are bosons, the quantum corrections 
always lead to an enhancement of the asymmetry compared to the results without
the corrections and the asymmetry $\eta$ has a maximum for moderate washout 
factors $\kappa\simeq \Ckappamaxeta$. The maximum of the relative enhancement 
of about $26\%$ is reached at $\kappa\simeq \Ckappacmaxreleta$.
The enhancement of $\epsilon_1$ due to the quantum corrections 
is suppressed at large washout factors, since the same processes which 
create and diminish the asymmetry, are effective at late times, where 
the \CP-violating parameter takes smaller values (compare
\fig\ref{epsilonepsilonvacvsmt}). 
In the opposite case of small $\kappa$, the particles decay late 
so that the washout is ineffective. However, the interval of integration 
in \eqn\eqref{LRhoMed} is located at relatively large momenta since the mass 
increasingly dominates the relativistic energies as the momenta are
red shifted to smaller values. This means that the integration is over an
interval in which the distribution $\f{\bbar}{}$ becomes smaller and smaller.
Therefore, the relative quantum corrections go to zero for small $\kappa$.

This interpretation of the results was already given in \cite{Garny:2009rv} 
as they are the same for the vertex and self-energy contributions in the 
hierarchical limit. We can here draw the additional conclusion that the 
combined effect from both contributions is the same in this limit. This 
is important since they could in principle have opposite effects. 
This is not the case as the self-energy and vertex \CP-violating 
parameters differ only by a positive prefactor, just as in the vacuum case. 
However, differences between the two can appear in the resonant regime. In 
this case the expressions for the vertex and self-energy contributions to 
$\epsilon_i$ have a different momentum dependence. Additionally, in general 
one has to take into account a further Boltzmann equation for the second 
heavy species $\psi_2$, which is currently being studied. We would like 
to stress again that  the size and sign of the corrections depend  on the 
quantum statistics of the particles in the vertex- and self-energy loops 
and will be different in a phenomenological scenario.

% =============================================================================
\section{\label{Summary}Conclusions and outlook}
% =============================================================================

In this paper, we have studied leptogenesis in a simple toy model 
consisting of one complex and two real scalar fields in a top-down 
approach,  using the Schwinger--Keldysh/Kadanoff--Baym formalism as 
starting point. This treatment, based on nonequilibrium quantum field 
theory techniques, is motivated by the fact that it allows a unified 
description of two key ingredients of leptogenesis, namely deviation 
from thermal equilibrium and loop-induced \CP-violation.

We find that the structure of the kinetic equations automatically 
ensures that no asymmetry is produced in thermal equilibrium.  In 
other words there is no need for the real intermediate state subtraction, 
i.e.~the formalism is free of the double-counting  problem typical 
for the canonical approach.

One of the key quantities in leptogenesis is the \CP-violating parameter.
Earlier studies have shown that there are two sources of \CP-violation, 
the vertex and the self-energy contribution. In this work, we have concentrated 
on the latter. We have found that for scalar fields  medium effects 
increase the self-energy contribution to the \CP-violating parameter. 

Contrary to the results obtained earlier in the framework of thermal field 
theory by replacing the zero temperature propagators with finite temperature 
propagators in the matrix elements of the Boltzmann equation, the medium 
corrections depend only linearly on the particle number densities.

Although the formal description of the self-energy and vertex contributions
to the \CP-violating parameter is technically quite different, the results 
for both are very similar qualitatively and, in the hierarchical case, even
quantitatively. In this work, we additionally studied the quasidegenerate 
case, for which the self-energy contribution is essential.

We have shown that the canonical expression for the self-energy
\CP-violating parameter is only applicable in the hierarchical case,
even though it does not diverge in the limit of equal masses. In the
resonant regime the interactions modify the mass spectrum of the 
quasiparticle excitations. Furthermore, using the Kadanoff--Baym 
formalism, it is possible to take a resummation of resonant contributions 
into account. Both effects lead to changes in the expression for the 
\CP-violating parameter. For moderate values of the degeneracy 
parameter $R$ the resonance corrections can enhance the
\CP-violating parameter by a factor of two. Therefore, it is 
important to take these corrections into account in  numerical simulations.

Another important effect is the resonant and medium enhancement of the total
decay widths. It leads to a faster decay of the heavy particles and more
efficient washout of the generated asymmetry. Therefore,  the increase
in the generated asymmetry due to the enhancement of the \CP-violating parameter
can be partially compensated by the increase of the in-medium decay widths.

In the ``maximal resonant'' regime the Boltzmann picture is no longer 
applicable. This can be attributed to the fact, that in this regime the 
peaks of the spectral functions of the heavy (toy-)neutrino fields 
overlap and it is no longer possible to unambiguously define  quasiparticles 
and one-particle distribution functions. Furthermore, the off-diagonal 
components of the correlation functions no longer have two pronounced 
peaks and therefore it is not possible to describe the mixing effects 
in terms of the corresponding on-shell \CP-violating 
parameters. Consequently, in this regime the calculation of the generated 
asymmetry requires us to use at least two-by-two matrix 
equations for the diagonal and off-diagonal components of the propagators
of the heavy fields with (in general) off-shell momenta. These can 
be obtained from the quantum kinetic equations by dropping the Poisson
brackets on the right-hand side. Furthermore, since the microscopic time scales
$(t_{mic} \sim \Delta M^{-1})$ and the macroscopic time scales $(t_{mac} \sim 
\Gamma^{-1})$ can be of the same order of magnitude in the maximal resonant 
regime,  memory effects can play an important role. Taking both off-shell
\emph{and} memory effects into account consistently requires the 
use of the full system of Kadanoff--Baym equations.

The formalism developed in this paper also provides  a powerful tool for 
analyzing quantum nonequilibrium effects induced by the expansion of the 
early Universe. In particular, there is a small additional 
``spontaneous'' contribution to the \CP-violation in the system similar to that 
encountered in  electroweak baryogenesis 
\cite{Konstandin:2004gy,Prokopec:2004ic,Prokopec:2003pj,Konstandin:2005cd}
This effect will be investigated in a forthcoming paper \cite{Garny:2009ab}.

\subsection*{Acknowledgements}
\noindent
This work was supported by the ``Sonderforschungsbereich'' TR27 and
by the ``cluster of excellence Origin and Structure of the Universe''.
We would like to thank J.\,Berges, J-S.\,Gagnon, A.\,Ibarra and
M.\,M.\,M{\"u}ller for helpful comments and discussions.

\begin{appendix}

% =============================================================================
\section{\label{cpclassic}CP-violating parameter}
% =============================================================================

In this appendix, we review the calculation of the self-energy contribution to
the \CP-violating parameter in vacuum,
$\epsilon_i^{\it vac} \equiv \left(\Gamma_{\psi_i \rightarrow{b}{b}}
-\Gamma_{\psi_i \rightarrow\bar{b}\bar{b}}\right)/
\left(\Gamma_{\psi_i \rightarrow{b}{b}}
+\Gamma_{\psi_i \rightarrow\bar{b}\bar{b}}\right)$, 
in the conventional in-out formalism.

Since the toy-Majoranas  are unstable, they cannot appear as in- or out-states
of S-matrix elements. Instead, their properties are defined by S-matrix
elements for scattering of stable particles mediated by the unstable neutrino
\cite{Veltman:1963th}. Resumming the propagator of the intermediate heavy
state, we can separate two-body scattering processes in resonance contributions 
and the rest. The \CP-violating part of the resonance contribution can then be 
interpreted as a characteristic of the on-shell intermediate toy-Majorana
\cite{PhysRevD.56.5431,Plumacher:1997ru}.
\begin{figure}[h!]
 \includegraphics[width=0.65\columnwidth]{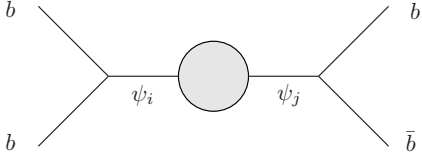}
\caption{\label{22scattering}Resummed diagrams contributing to the resonant part
of the 2$\rightarrow$2 scattering amplitude of the process $bb\rightarrow \bar
b\bar b$.}
\end{figure} 
The  amplitude of the $s$-channel two-body scattering process $bb\rightarrow
\bar b
\bar b$ (see Fig.\,\ref{22scattering}) can conveniently be expressed as 
\begin{align}
\label{ScatteringVac}
\mathcal{M}_{bb\rightarrow \bar b\bar b}=\sum_{i,j}\Gamma_i^A
G^{ij}(s)\Gamma_j^B\,,
\end{align}
where $\Gamma_i^A$ and $\Gamma_j^B$ represent the vertices $\psi_i bb$ and 
$\psi_j \bar b\bar b$ that include the wave functions of the initial and 
final states, and $G^{ij}$ are the full propagators obtained by resumming an 
infinite series of toy-Majorana self-energy graphs \cite{PhysRevD.56.5431}.

The resummation can be performed using the Schwinger-Dyson equation in vacuum:
\begin{align}
\label{SDVac}
[G^{-1}]^{ij}(p^2)=\left[p^2-M_i^2\right]\delta^{ij}-\Pi^{ij}(p^2)\,.
\end{align}
At one-loop level the self-energy $\Pi^{ij}$ reads
\begin{align}
\label{SelEnReg}
\Pi^{ij}(p^2)=-\frac{g_i g_j^*+g_i^* g_j}{32\pi^2}B(p^2)\,,
\end{align}
where 
\begin{align}
B(p^2)=\Delta +2-\ln \frac{|p^2|}{\mu^2} +i\pi\theta(p^2)\,,
\end{align}
and $\Delta\equiv \epsilon^{-1}-\gamma+4\pi$ contains the divergent 
contribution. 
To renormalize the mass and the self-energy we introduce the wave-function and
mass counterterms to the Lagrangian: 
\begin{align}\label{lagrangianRenormVac}
   \delta{\cal L} & = \frac12 \delta Z_{ij} \partial^\mu \psi_i \partial_\mu
\psi_j  -\frac12 \delta M^2_{ij} \psi_i\psi_j \,,
\end{align}
where $\delta Z_{ij}$ and $\delta M^2_{ij}$ are symmetric two-by-two matrices.
This 
implies that the renormalized self-energy is given by 
\begin{align}
\label{PiRenVac}
\Pi^{ij}_{ren}(p^2)=\Pi^{ij}(p^2)-p^2\delta Z_{ij}+\delta M^2_{ij}\,.
\end{align}
In the on-shell renormalization scheme the \textit{dispersive} parts of the 
components of the renormalized self-energy must satisfy the conditions 
\begin{subequations}
\label{RenCondVac}
\begin{align}
\label{DiagRenCondVac}
\Pi_{ren}^{ii}(p^2=M_i^2) & = \frac{d}{dp^2} \Pi_{ren}^{ii}(p^2=M_i^2)  =  0\,,
\\
\label{OffDiagRenCondVac}
 \Pi_{ren}^{ij}(p^2=M_i^2) &  =  \Pi_{ren}^{ij}(p^2=M_j^2)  =  0\quad (i\neq
j)\,.
\end{align}
\end{subequations} 
Using \eqref{PiRenVac} and \eqref{RenCondVac} and the
explicit form of the bare self-energy   we 
calculate $\delta Z_{ij}$ and $\delta M^2_{ij}$. Substituting them into 
\eqref{PiRenVac} we find
\begin{subequations}
\label{PiRenVacExpl}
\begin{align}
\Pi_{ren}^{ii} &=\frac{|g_i|^2}{16\pi^2}\left[
\ln \frac{|p^2|}{M_i^2} -\frac{p^2-M_i^2}{M_i^2}-i\pi\theta(p^2)
\right]\,,\\
\Pi_{ren}^{ij}&=\frac{\Re(g_ig_j^*)}{16\pi^2}\left[
\frac{p^2-M_i^2}{M_j^2-M_i^2}\ln\frac{|p^2|}{M_j^2}\right.\nonumber\\
&\hspace{1.2cm}+\left.\frac{p^2-M_j^2}{M_i^2-M_j^2}\ln \frac{|p^2|}{M_i^2}
-i\pi\theta(p^2)
\right]\,.
\end{align}
\end{subequations}
Inverting \eqref{SDVac} we obtain for the components of the renormalized 
resummed propagator:
\begin{subequations}
\label{GijRenVac}
\begin{align}
G^{ii}(p^2)&=+[G^{-1}]^{jj}(p^2)/\det [G^{-1}(p^2)]\,,\\
G^{ij}(p^2)&=-[G^{-1}]^{ij}(p^2)/\det [G^{-1}(p^2)]\,.
\end{align}
\end{subequations}
Because of the presence of   absorptive terms in \eqref{PiRenVacExpl} the 
determinant of the inverse propagator in \eqref{GijRenVac} has 
two poles in the complex plane at 
\begin{align}
s_i\simeq M_i^2-iM_i\Gamma_i\,,
\end{align}
where $\Gamma_i=|g_i|^2/16\pi M_i$ is the tree-level decay 
width of $\psi_i$. Expanding \eqref{GijRenVac} around the poles 
and substituting the leading expansion terms to \eqref{ScatteringVac}
we find \cite{Pilaftsis:2003gt}
\begin{align}
\mathcal{M}_{bb\rightarrow \bar b \bar b}\simeq 
\sum_i V_i^A(s)\frac{1}{s-s_i}V_i^B(s)\,,
\end{align}
where 
\begin{align}
\label{ViAB}
V_i^{A(B)}(s)&\equiv\Gamma_i^{A(B)}-\frac{[G^{-1}]^{ij}(s)}{[G^{-1}]^{jj}(s)}
\nonumber\\
&=\Gamma_i^{A(B)}+\frac{\Pi^{ij}(s)}{s-M_j^2-\Pi^{jj}(s)}
\Gamma_j^{A(B)}\,.
\end{align}
The modulo squared of the $bb\rightarrow \bar b \bar b$ scattering amplitude is
then given by 
\begin{align}
\label{MbbbbSqVac}
|\mathcal{M}_{bb\rightarrow \bar b \bar b}|^2&\simeq 
\sum_i
\frac{|V_i^A(s)|^2|V_i^B(s)|^2}{M_i\Gamma_i}\frac{M_i\Gamma_i}{
(s-M_i)^2+(M_i\Gamma_i)^2}\nonumber\\
&+ {\rm cross~terms.}
\end{align}
The Breit-Wigner propagators in the diagonal terms of \eqref{MbbbbSqVac} 
strongly peak on the mass shell of the quasiparticles, i.e.~at $s=M_i^2$ 
and rapidly decrease off the mass shell. In the limit of vanishing decay 
widths 
\begin{align}
\label{BreitWignDeltaLim}
\frac{M_i\Gamma_i}{(s-M_i)^2+(M_i\Gamma_i)^2}\rightarrow 
\pi\,\delta(s-M_i^2)\,.
\end{align}
Furthermore, if  $|M_i^2-M_j^2|\gg M_i\Gamma_i,M_j\Gamma_j$ then the 
two Breit-Wigner propagators do not overlap and we can neglect 
the cross terms in \eqref{MbbbbSqVac}. In other words, in this 
approximation the resonant (real intermediate state) part of 
the amplitude is given by
\begin{align}
\label{MbbbbSqVacDeltaLim}
|\mathcal{M}_{bb\rightarrow \bar b \bar b}|^2\simeq 
\sum_i \frac{|V_i^A(s)|^2|V_i^B(s)|^2}{M_i\Gamma_i}\,\pi\,\delta(s-M_i^2)\,.
\end{align}
Equation \eqref{MbbbbSqVacDeltaLim} suggests, that $V_i^{A(B)}$
should also be evaluated at $s=M_i^2$. As follows from  \eqref{ViAB}
to leading order in the couplings its modulo squared can be represented 
in the form
\begin{align}
|V_i^{A(B)}|^2\simeq \Gamma_i^2(1-\epsilon_i)\,.
\end{align}
Using   the explicit form of the renormalized self-energy
\eqref{PiRenVacExpl} we obtain for the \CP-violating parameters
$\epsilon_i$ in vacuum:
\begin{align}
   \label{epsilonClassVac}
   \epsilon_i = -\frac{|g_j|^2}{16\pi}
   {\rm Im}\biggl(\frac{g_i g_j^*}{g_i^* g_j}\biggr)
   \frac{M_j^2-M_i^2}{(M_j^2 -M_i^2)^2+M_j^2 \Gamma_j^2}\,.
\end{align}
Because of the presence of the $M_j\Gamma_j$ term in the denominator of
\eqref{epsilonClassVac}, it does not diverge if $M_i\rightarrow M_j$. 
However, the condition $|M_i^2-M_j^2|\gg M_i\Gamma_i,M_j\Gamma_j$ 
is not satisfied in this case and therefore the approximations we have made 
to derive  \eqref{epsilonClassVac} are not valid in this limit.
Let us also note that due to approximation \eqref{BreitWignDeltaLim}
the \CP-violating parameter \eqref{epsilonClassVac} characterizes 
\textit{on-shell} heavy particles. 
 
Integrals of the left- and right-hand sides of \eqref{BreitWignDeltaLim}
over $s$ are equal only if we integrate  in the range from $-\infty$ to 
$+\infty$. But in the $s-$channel the momentum transfer squared is always 
positive. Moreover, the functions $V_i^{A(B)}$ also depend on $s$. Thus,
in the transition from \eqref{MbbbbSqVac} to \eqref{MbbbbSqVacDeltaLim},
i.e.~in 
replacing the Breit-Wigner propagator by a Dirac $\delta$-function and
evaluating
$V_i^{A(B)}$ at $s=M_i^2$ we have made an approximation which, strictly 
speaking, is only valid in the limit $\Gamma_i\rightarrow 0$.
We could perform the integration more carefully and take the 
finite widths $\Gamma_i$ into account using  Cauchy's integral 
theorem and evaluating $V_i^{A(B)}$ at the poles $s_i=M_i^2-iM_i\Gamma_i$. This
leads to a slightly different expression for the 
\CP-violating parameter \cite{Anisimov:2005hr}:
\begin{align}
   \label{epsilonPlumVac}
   \epsilon_i = -\frac{|g_j|^2}{16\pi}
   {\rm Im}\biggl(\frac{g_i g_j^*}{g_i^* g_j}\biggr)
   \frac{M_j^2-M_i^2}{(M_j^2 -M_i^2)^2+(M_j \Gamma_j-M_i \Gamma_i)^2}\,.
\end{align}
Equation \eqref{epsilonPlumVac} can be considered as a better estimate 
of the \CP-violating effects in the system. However, since the
\textit{on-shell} 
approximation \eqref{BreitWignDeltaLim} no longer applies, one can not interpret
\eqref{epsilonPlumVac} as an expression for the \CP-violating parameter of an 
\textit{on-shell} state. In other words, strictly speaking,
\eqref{epsilonPlumVac} would require us to use an \textit{off-shell}
generalization of the Boltzmann equation.

Let us also note that in the case  $|M_i^2-M_j^2|\gg M_i\Gamma_i,M_j\Gamma_j$,
which 
we have considered here, the difference between the two expressions for the 
\CP-violating parameter can be neglected.

% =============================================================================
\section{\label{real}Mixing real scalar fields}
% =============================================================================

In this appendix, we derive the Kadanoff--Baym, quantum kinetic
and Boltzmann equations for a system of two
(or, in general, $n$) mixing real scalar fields.

\subsection{\label{SD}Schwinger--Dyson equation}

\begin{figure}[th!]
   \includegraphics[width=0.9\columnwidth]{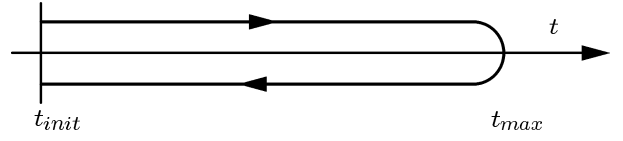}
   \caption{\label{contour}Closed real-time path $\cal C$.}
\end{figure}

The  generating functional of such a system reads
\begin{align}
   \label{genfunct}
   {\cal Z} = \intg\mathscr{D}\psi_1\mathscr{D}\psi_2
   \exp[i(S+J_i \psi_i+{\textstyle\frac12}\psi_i K_{ij}\psi_j)]\,.
\end{align}
The bilinear external source  is a $n$-by-$n$ matrix with 
the property $K_{ij}(x,y)=K_{ji}(y,x)$, where $n=2$ in the toy-model. Note that
the field and the external sources are defined on the positive and negative 
branches of a closed real-time contour; see Fig.\,\ref{contour}.
Throughout this work, we use the compact notation of
Ref.~\cite{Danielewicz:1982kk}, which avoids the doubling of the degrees of
freedom. In particular, note that the indices $i,j\in\{1,\dots,n\}$ refer to
the two real scalar fields $\psi_i$ in the toy model.

The functional derivatives of the generating functional for connected Green's
functions, ${\cal W}=-i\ln {\cal Z}$, with respect to the external sources read
\begin{subequations}
   \begin{align}
      \frac{\partial {\cal W}}{\partial J_i(x)}      & = \Psi_i(x)\,, \\
      \frac{\partial {\cal W}}{\partial K_{ij}(x,y)} & =
      {\textstyle\frac12}[G_{ji}(y,x)+\Psi_i(x)\Psi_j(y)]\,,
   \end{align}
\end{subequations}
where $\Psi_i(x)\equiv \langle \hat\psi_i(x)\rangle$ is the 
expectation value of the field operator.
The propagator $G_{ji}$ is  a $n$--by--$n$ matrix; its off-diagonal components
describe the mixing of the two fields. We emphasize again that the indices
refer to the field content, and have nothing to do with the branches of the
time-contour.

Performing a Legendre transform of the generating functional for connected
Green's functions, we obtain the effective action
\begin{align}
   \Gamma\equiv {\cal W} -
   J_i\Psi_i-{\textstyle\frac12}\tr[K_{ij}G_{ji}]
   -{\textstyle\frac12}\Psi_i K_{ij} \Psi_j\,.
\end{align}
Its functional derivatives with respect to the expectation value and the
propagator reproduce the external sources:
\begin{subequations}
   \label{DiffGamma_real}
   \begin{align}
      \frac{\delta \Gamma}{\delta \Psi_i(x)}&=-J_i(x)
      -\intg\dg z~ K_{ij}(x,z)\Psi_j(z)\,,\\
      \hskip -1mm
      \frac{\delta \Gamma}{\delta G_{ij}(x,y)}&=
      -{\textstyle\frac12}K_{ji}(y,x)\,,
   \end{align}
\end{subequations}
where $\dg z \equiv \sqrt{-g}\,d^4z$ and $g$ is the determinant of the
space-time metric.

Shifting the fields by their expectation values
$\psi_i\rightarrow\psi_i+\Psi_i$,
we can rewrite the effective action in the form
\begin{align}
   \Gamma = & -i\ln \intg \mathscr{D}\psi_1\mathscr{D}\psi_2
              \exp [i(S+J_i\psi_i+{\textstyle\frac12}\psi_i K_{ij}\psi_j)]
              \nonumber\\
            & + S_{cl}[\Psi]-{\textstyle\frac12}\tr[K_{ij}G_{ji}]\,.
\end{align}
Next, we tentatively write the effective action in the form
\begin{align}
   \label{G2deff}
   \Gamma\equiv S_{cl}[\Psi]+{\textstyle\frac{i}{2}}
\ln\det\left[G^{-1}\right]+{\textstyle\frac12}\tr[\mathscr{G}_{ij}^{-1}G_{ji}]
   +\Gamma_2\,,
\end{align}
which defines the functional $\Gamma_2$. 

Differentiation of the third term on the right-hand side with 
respect to the field propagators yields the inverse free propagators.
Note that the latter one  is diagonal -- in the absence of interactions 
the two fields do not mix:
\begin{align}
   \mathscr{G}^{-1}_{ij}(x,y)=i(\square_x+M^2_i)\,\delta^g(x,y)\delta_{ij} \,,
\end{align}
where $\delta^g(x,y)=(-g_x)^{-\frac14}\delta(x,y)(-g_y)^{-\frac14}$
is a generalized Dirac $\delta$-function.

The second term on the right-hand side is defined by 
the path integral 
\begin{align*}
   \det\left[\frac{G^{-1}}{2\pi}\right]\equiv\int 
   \mathscr{D}\psi_1\mathscr{D}\psi_2
   \exp\left(\psi_i\, {G^{-1}_{ij}} \psi_j\right)\,.
\end{align*}
To calculate the functional derivative of $\ln\det\left[G^{-1}\right]$,
we  take into account that 
\begin{align}
   \label{inverseprop}
   \intg \dg z\, G^{-1}_{mk}(u,z)G_{kn}(z,v)=\delta^g(u,v)\,\delta_{mn}~.
\end{align}
After some algebra and use of \eqref{inverseprop}, we obtain  
\begin{align}
   \frac{\delta}{\delta G_{ij}(x,y)}\ln\det \left[G^{-1}\right]=
   -G^{-1}_{ji}(y,x)\,.
\end{align}
The functional derivative of \eqref{G2deff} with respect to $G$ then reads
\begin{align}
   \label{G2diff}
   \frac{\delta \Gamma }{\delta G_{ij}(x,y)}
   = & -{\textstyle\frac{i}{2}}G^{-1}_{ji}(y,x)
       +{\textstyle\frac{i}2}\mathscr{G}_{ji}^{-1}(y,x)
       +\frac{\delta \Gamma_2}{\delta G_{ij}(x,y)}\nonumber\\
   = & -{\textstyle\frac12}K_{ji}(y,x)\,.
\end{align}
The considered physical situation corresponds to vanishing sources. Introducing 
the self-energy, 
\begin{align}
   \label{self-energydeffG}
   \Pi_{ij}(x,y)\equiv 2i\frac{\delta \Gamma_2}{\delta G_{ji}(y,x)} \,,
\end{align}
we can then rewrite \eqref{G2diff} in the form
\begin{align}
   \label{SDeqs}
   G^{-1}_{ij}(x,y)=\mathscr{G}_{{ij}}^{-1}(x,y)-\Pi_{ij}(x,y)\,.
\end{align}
As can be inferred from \eqref{SDeqs}, off-diagonal components 
of $G_{ij}$ are induced by off-diagonal components of $\Pi_{ij}$. 
In the considered model they arise because both real scalars couple 
to the ``baryons''.

\subsection{Kadanoff--Baym equations}

Convolving the Schwinger--Dyson equations \eqref{SDeqs} with $G$ from
the right and using \eqref{inverseprop}, we obtain
\begin{align}
   \label{convolution}
   i\left[\square_x+M_i^2\right] & G_{ij}(x,y) =
   \delta^g(x,y)\delta_{ij}\nonumber\\
   &+\intg\dg z\,\Pi_{ik}(x,z)G_{kj}(z,y)\,.
\end{align}

The statistical propagators and spectral functions are defined by
\begin{subequations}
   \label{GFGrhoij}
   \begin{align}
      G_{F}^{ij}(x,y)    & = {\textstyle\frac12}\langle 
                           [\psi_i(x),\psi_j(y)]_{+} \rangle\,,\\
      G_{\rho}^{ij}(x,y) & = i\langle [\psi_i(x),\psi_j(y)]_{-} \rangle\,.
   \end{align}
\end{subequations}
From the definitions \eqref{GFGrhoij} it follows that
\begin{align}
   \label{GFGrhoProp}
   G_{F}^{ij}(x,y)=G_{F}^{ij}(y,x),\,
   G_{\rho}^{ij}(x,y)=-G_{\rho}^{ij}(y,x)\,.
\end{align}
The time-ordered Schwinger--Keldysh propagator (which is the analogue 
of the Feynman propagator on the closed time path) is a linear 
combination of these functions:
\begin{align}
   \label{GfGr}
   \hspace{-1mm}
   G_{ij}(x,y)=G_{F}^{ij}(x,y)-{\textstyle\frac{i}{2}}\sign_{\cal C}(x^0-y^0)
   G_{\rho}^{ij}(x,y).
\end{align}

Upon use of the signum- and $\delta$-function differentiation 
rules, the action of the $\square_x$ operator on the second term 
in \eqref{GfGr} yields a product of $g^{00}\delta(x^0,y^0)$ and 
$\nabla^x_0 G_{\rho_{ij}}(x,y)$. Using the definition \eqref{GFGrhoij} and 
the canonical commutation relations \cite{Isham:1978}
\begin{align}
   \lim\limits_{y^0\rightarrow
x^0}[\psi_i(x^0,\vec{x}\,),\pi_j(x^0,\vec{y}\,)]_{-}&=
   i\delta (\vec{x},\vec{y})\delta_{ij},
\end{align}
where $\pi_j=g^{00}\sqrt{-g}\,\nabla_0\psi_j$\,, we find for the derivative of
the spectral function
\begin{align}
   \label{Grhoder}
   \nabla^x_0G_{\rho}^{ij}(x,y)=
   \frac{\delta(\vec{x},\vec{y}\,)\delta^{ij}}{g^{00}\sqrt{-g}}.
\end{align}
Multiplication of \eqref{Grhoder} by $g^{00}\delta(x^0,y^0)$ 
then  gives the generalized Dirac $\delta$-function $\delta^g(x,y)\delta^{ij}$,
which cancels the $\delta$-function on the right-hand side of 
\eqref{convolution}. Furthermore, we decompose the self-energy according to
\begin{align}
   \hspace{-1mm}
   \label{PifPir}
   \Pi_{ij}(x,y)=\Pi_{F}^{ij}(x,y)-{\textstyle\frac{i}{2}}
   \sign_{\cal C}(x^0-y^0)\Pi_{\rho}^{ij}(x,y) \;.
\end{align}
The resulting system of Kadanoff--Baym equations reads (see
\cite{Berges:2004yj,Hohenegger:2008a} for more details):
\begin{subequations}
   \label{KBeqs_real}
   \begin{align}
      \hspace{-1mm}
      \label{GFeq}
      [\square_x+M_i^2]G_{F}^{ij}(x,y) & =
      {\textstyle\int\limits^{y^0}_0}\dg
z\,\Pi_{F}^{ik}(x,z)G_{\rho}^{kj}(z,y)\nonumber\\
      & - {\textstyle\int\limits^{x^0}_0}\dg
z\,\Pi_{\rho}^{ik}(x,z)G_{F}^{kj}(z,y)\,,\\
      \hspace{-1mm}
      \label{Grhoeq}
      [\square_x+M_i^2]G_{\rho}^{ij}(x,y) & =
      {\textstyle\int\limits_{x^0}^{y^0}}\dg
z\,\Pi_{\rho}^{ik}(x,z)G_{\rho}^{kj}(z,y)\,.
   \end{align}
\end{subequations}
In the limit of just one scalar field it reverts to the system derived in
\cite{Hohenegger:2008zk}. Numerical solutions of full Kadanoff--Baym equations, 
for systems involving effectively a single degree of freedom in the scalar
sector, 
have been studied e.g. in~\cite{Danielewicz:1982ca,Berges:2000ur,Berges:2001fi,
Aarts:2001yn,Berges:2002wr,Juchem:2003bi,Arrizabalaga:2005tf,Lindner:2005kv,
Lindner:2007am}.

\subsection{Quantum kinetics}

The system of  Kadanoff--Baym equations can be rewritten in terms of the
advanced and retarded propagators, $G_{R}^{ij}$ and $G_{A}^{ij}$:
\begin{align}
   \label{KBQK}
   & [\square_x+M_i^2]G_{F(\rho)}^{ij}(x,y)=-\intg\dg z
     \theta(z^0)\nonumber\\
   & \times[\Pi_{F(\rho)}^{ik}(x,z)G_{A}^{kj}(z,y)+
     \Pi_{R}^{ik}(x,z)G_{F(\rho)}^{kj}(z,y)]\,.
\end{align}
In order to close the system, Eqs. \eqref{KBQK} must be supplemented
with the analogous equations for the retarded and advanced propagators:
\begin{align}
   \label{GRAQK}
   [\square_x+M_i^2] & G_{R(A)}^{ij}(x,y)=\delta^g(x,y)\delta^{ij}\nonumber\\
                     & -\intg \dg z\, \Pi_{R(A)}^{ik}(x,z)G_{R(A)}^{kj}(z,y)\,.
\end{align}

From the definitions of the retarded and advanced propagators and relations
\eqref{GFGrhoProp} one can infer that 
\begin{align}
   \label{GRGARel}
   G_{R}^{ij}(x,y) & \equiv \theta(x^0-y^0)G_{\rho}^{ij}(x,y)\nonumber\\
                   & = -\theta(x^0-y^0)G_{\rho}^{ji}(y,x)=G_{A}^{ji}(y,x)\,.
\end{align}
Therefore, after interchanging $x$ and $y$ and then $i$ and $j$ in
\eqref{KBQK}, we obtain 
\begin{align}
   \label{GFQKichng}
   [&\square_y+M_j^2]G_{F(\rho)}^{ij}(x,y)=-\intg\dg z
   \theta(z^0)\nonumber\\
   &\times[G_{R}^{ik}(x,z)\Pi_{F(\rho)}^{kj}(z,y)+
   G_{F(\rho)}^{ik}(x,z)\Pi_{A}^{kj}(z,y)]\,.
\end{align}
The same operation applied to \eqref{GRAQK} yields
\begin{align}
   \label{GRAQKintch}
   [\square_y+M_j^2]&G_{A(R)_{ij}}(x,y)=\delta^g(x,y)\delta_{ij}\nonumber\\
   &-\intg \dg z\,G_{A(R)_{ik}}(x,z) \Pi_{A(R)_{kj}}(z,y)\,.
\end{align}

Following the usual procedure, we introduce the center and relative 
coordinates $X$ and $s$ \cite{PhysRevD.32.1871}. The Wigner transform of 
the statistical propagator is defined by
\begin{align}
   \label{WignerTrafo}
   G^{ij}_F(X,p)&=\sqrt{-g}_X\intg d^4s\, e^{ips}\,G^{ij}_F(X,s)\,.
\end{align}
The definition of the Wigner transform of $G^{ij}_\rho(X,s)$ differs from
\eqref{WignerTrafo} by an overall factor of $-i$.  From \eqref{GFGrhoProp} it
then follows that 
\begin{align}
   \hspace{-1mm}
   \label{GFGrhoComplConj}
   \hat{G}^{*}_{F}(X,p)=\hat{G}^{T}_{F}(X,p),\,\, 
   \hat{G}^{*}_{\rho}(X,p)=\hat{G}^{T}_{\rho}(X,p)\,,
\end{align}
where the hats denote the corresponding matrices and the superscript `T'
denotes transposition. As we see, the Wigner transforms are Hermitian matrices.
The Wigner transforms of the retarded and advanced propagators are defined
analogously to \eqref{WignerTrafo}. Using \eqref{GRGARel} one can then show
that 
\begin{align}
   \label{GRGAconj}
   \hat{G}_{R}(X,p)=\hat{G}^{\dagger}_{A}(X,p)\,.
\end{align}
Just as in the case of a single real scalar field \cite{Hohenegger:2008zk}, 
from the definitions of the Wigner transforms, relation \eqref{GRGARel} and 
the equality  $\theta(s^0)+\theta(-s^0)=1$ it follows that 
\begin{align}
   \label{GRminGA}
   \hat{G}_{R}(X,p)-\hat{G}_{A}(X,p)=i\hat{G}_{\rho}(X,p)\,.
\end{align}
We could have derived Eqs.\,\eqref{GRGAconj} and \eqref{GRminGA} 
using the spectral representation of the retarded and advanced 
propagators
\begin{align}
   \hat{G}_{R(A)}(X,p) = -\int \frac{d\omega}{2\pi}
   \frac{\hat{G}_\rho(X,\omega,\vec{p}\,)}{p_0-\omega\pm i\epsilon}\,.
\end{align}
The spectral representation also implies that $\hat G_R$ and $\hat G_A$ can 
be represented as a linear combination of two Hermitian matrices:
\begin{align}
   \label{GRGAdecomposition}
   \hat{G}_{R(A)}(X,p)=\hat{G}_{h}(X,p)\pm   
   {\textstyle\frac{i}{2}}\,\hat{G}_{\rho}(X,p)\,.
\end{align}
Let us now subtract \eqref{GFQKichng} from \eqref{KBQK} and Wigner trans\-form
the left- and the right-hand sides of the resulting equation. On the
right-hand 
side we neglect the memory effects, that is, we discard the  $\theta$ function 
and replace $X_{xz},X_{zy}$  by $X_{xy}$.  Furthermore, we keep only  terms 
at most linear in the Wigner transform of the covariant derivative 
${\cal D}_\alpha$; see \cite{Hohenegger:2008zk} for more details. Introducing 

the Poisson brackets,
\begin{align}
\label{PoissonBrackets}
\{{A}(X,p),{B}(X,p)&\}_{PB}\equiv
\frac{\partial}{\partial p_\alpha} {A}(X,p) {\cal D}_\alpha{B}(X,p)\nonumber\\
&-{\cal D}_\alpha {A}(X,p) \frac{\partial}{\partial p_\alpha}{B}(X,p)\,,
\end{align}
the generalized Poisson brackets\footnote{If the matrices $\hat A(X,p)$ and 
$\hat B(X,p)$ are hermitian 
and commute, then $[A(X,p),B(X,p)]_{PB}=\{A(X,p),B(X,p)\}_{PB}$.}
\begin{align}
   \label{GeneralizedPB}
   \hspace{-1mm}
   [\hat A&(X,p), \hat B(X,p)]_{PB}\equiv -i[\hat A(X,p),\hat B(X,p)]_-\\
   &+{\textstyle\frac12}\{\hat A(X,p),\hat B(X,p)\}_{PB} 
   +{\textstyle\frac12}\{\hat A(X,p),\hat B(X,p)\}^\dagger_{PB}
   \,,\nonumber
\end{align}
and 
\begin{align}
\label{OmegaRADef}
\hat \Omega_{R(A)}(X,p) \equiv \hat p^2-\hat M^2-\hat \Pi_{R(A)}(X,p) 
\end{align}

we can rewrite the result in a compact form:
\begin{align}
   \label{GFrhoQKeq}
   \bigl[\hat\Omega_R&(X,p),\hat G_{F(\rho)}(X,p)\bigr] _{PB}\nonumber\\
   &=\hat G_{F(\rho)}(X,p) \hat \Pi_\rho(X,p)-\hat \Pi_{F(\rho)}(X,p)
   \hat G_\rho(X,p)\nonumber\\
   &-\bigl[\hat G_R(X,p),\hat \Pi_{F(\rho)}(X,p)\bigr] _{PB}\,.
\end{align}
In the case of a single scalar field the commutator in \eqref{GeneralizedPB}
vanishes and, using \eqref{GRGAdecomposition}, we can then show that
the quantum kinetic equation \eqref{GFrhoQKeq} for the spectral
function and statistical propagator revert to those derived in 
\cite{Hohenegger:2008zk}. After some algebra we  rewrite 
the quantum kinetic equation for the spectral function in the form:
\begin{align}
   \label{GrhoQKeq}
   \hspace{-1mm}
   [\hat\Omega(X,p),\hat G_\rho(X,p)&]_{PB}\nonumber\\
   &=-[\hat G_h(X,p),\hat\Pi_\rho(X,p)] _{PB}\,,
\end{align}
where 
\begin{align}
\label{OmegaDef}
\hat \Omega \equiv \hat p^2-\hat M^2-\hat \Pi_h\,.
\end{align}
The quantum kinetic equations must be supplemented by the corresponding
constraint equations. To derive these, we add up the Wigner-transforms of 
\eqref{GFQKichng} and \eqref{KBQK}. To linear order in the gradients 
the resulting constraint equations read
\begin{align}
   \label{GFrhoconstraint}
   [\hat \Omega_R(X,p),&\hat G_{F(\rho)}(X,p)]_\oplus=[\hat
\Pi_{F(\rho)}(X,p),\hat G_A(X,p)]_\oplus\nonumber\\
   &+{\textstyle\frac{i}{2}}\{\hat \Pi_{F(\rho)}(X,p),\hat G_A(X,p)\}_{PB
\oplus}\nonumber\\
   &+{\textstyle\frac{i}{2}}\{\hat \Pi_R(X,p),\hat
G_{F(\rho)}(X,p)\}_{PB\oplus}\,,
\end{align}
where we have introduced $[\hat A,\hat B]_\oplus\equiv \hat A\hat B+
\hat B^\dagger \hat A^\dagger$ and 
$\{\hat A,\hat B\}_{PB\oplus}\equiv \{\hat A,\hat B\}_{PB}+
\{\hat B^\dagger, \hat A^\dagger\}_{PB}$ to shorten the notation.
In the case of a single scalar field \eqref{GFrhoconstraint} revert 
to those derived in \cite{Hohenegger:2008zk}. After some algebra 
we can again simplify the constraint equation for the spectral function:
\begin{align}
   \label{GrhoConstraint}
   [\hat \Omega(X,p),\hat G_\rho&(X,p)]_\oplus=[\hat \Pi_\rho(X,p),\hat
G_A(X,p)]_\oplus\nonumber\\
   &+{\textstyle\frac{i}{2}}\{\hat \Pi_\rho(X,p),\hat G_h(X,p)\}_{PB
\oplus}\nonumber\\
   &+{\textstyle\frac{i}{2}}\{\hat \Pi_h(X,p),\hat G_\rho(X,p)\}_{PB\oplus}\,.
\end{align}
To close the system of the constraint equations \eqref{GrhoConstraint}
we have to derive the constraint equation for the function $\hat G_h$ or, 
alternatively, for the retarded propagator:
\begin{align}
   \label{GRconstraint}
   [\hat \Omega_R(X,p),\hat G_R&(X,p)]_+=-2\hat I \nonumber\\
   &+{\textstyle\frac12}\{\hat \Pi_R(X,p), \hat G_R(X,p)\}_{PB+}\,,
\end{align}
where the $+$ operation is defined as the $\oplus$ 
operation but does not involve Hermitian conjugation. 
Unlike in the case of a single scalar field, the constraint equations 
\eqref{GrhoConstraint} and \eqref{GRconstraint} are not algebraic and 
cannot be solved easily. The solution of the \textit{algebraic} part 
of \eqref{GrhoConstraint} and \eqref{GRconstraint} is not a solution of 
the complete differential equations. Therefore, we conclude that even 
to linear order in the gradients the spectrum of the mixing scalar 
fields depends on the derivative terms described by the
Poisson brackets.

\subsection{Solution in the equilibrium limit}

Since in vacuum and in thermal equilibrium the system is homogeneous,
isotropic and static  the two-point functions are translationally 
invariant, i.e.~do not depend on the center coordinate $X$. Therefore, 
in thermal equilibrium and in vacuum, the Poisson brackets
\eqref{PoissonBrackets} vanish. 
In the following, we discuss the solutions of the constraint equations 
obtained by neglecting the Poisson brackets. However, we do \emph{not} 
make use of any relations relying on periodic equilibrium boundary 
conditions. Therefore, the formal solutions should also be a reasonable 
approximation sufficiently close to equilibrium, when gradient 
contributions are small. In this case, \eqref{GRconstraint} reduces to an
\emph{algebraic} matrix equation. Its solution reads
\begin{align}
   \label{GRSol}
   \hat G_R(X,p)=-\hat \Omega_R^{-1}(X,p)\,.
\end{align}
From  the definition \eqref{OmegaRADef}, 
relation  \eqref{GRGAconj} and the 
analogous relations for the retarded and advanced self-energies 
$\hat \Pi_{A(R)}$ it then follows that 
\begin{align}
   \label{GASol}
   \hat G_A(X,p)=-\hat \Omega_A^{-1}(X,p)\,.
\end{align}

From  relation \eqref{GRGAdecomposition} it follows that
\begin{align}
   \hat G_\rho(X,p) & = i[-\hat G_R(X,p)+\hat G_A(X,p)]\,.
\end{align}
Using \eqref{GRSol} and \eqref{GASol} we then find that in equilibrium 
and in vacuum

\begin{align}
   \label{GrhoSol}
   \hat G_\rho(X,p) & = i[\hat \Omega_R^{-1}(X,p)-\hat
\Omega_A^{-1}(X,p)]\nonumber\\
                    & =  -\hat \Omega_{R(A)}^{-1}(X,p)\hat \Pi_\rho(X,p)\hat
\Omega_{A(R)}^{-1}(X,p)\,.
\end{align}
Similarly to \eqref{GrhoSol}, the equilibrium (or  vacuum) solution 
of the constraint equation for the statistical propagator is given by
\begin{align}
   \label{GFsolution}
   \hat G_F(X,p) = -\hat\Omega^{-1}_R(X,p) \hat \Pi_F (X,p)
\hat\Omega^{-1}_A(X,p)\,.
\end{align}
Note  that unlike in \eqref{GrhoSol} the order of the retarded 
and advanced components is no longer arbitrary. Let us now split 
$\hat \Omega_{R(A)}$ into  off-diagonal and diagonal matrices, 
the latter being denoted by $\hat \varOmega_{R(A)}$. Next we 

 define \textit{diagonal} matrices 
\begin{align}
   \label{GRdiagSol}
   \hat {\cal G}_{R(A)}(X,p)&\equiv -\hat \varOmega_{R(A)}^{-1}(X,p)\
\end{align}
and
\begin{align}
   \hspace*{-2mm}
   \label{GFrhodiagSol1}
    \hat {\cal G}_{F(\rho)}(X,p)&\equiv -\hat \varOmega^{-1}_R(X,p)
    \hat\varPi_{F(\rho)} (X,p) \hat \varOmega^{-1}_A(X,p)\,,
\end{align}
where $\hat\varPi_{F(\rho)}$ denotes the diagonal components of 
the self-energy matrices $\hat\Pi_{F(\rho)}$.
Explicitly 
\begin{align}
   \label{GFrhodiagSol}
   {\cal G}^{ii}_{F(\rho)}(X,p)&= -\frac{\Pi^{ii}_{F(\rho)}(X,p)}
   {[\Omega^{ii}(X,p)]^2+{\textstyle\frac14}[\Pi^{ii}_\rho(X,p)]^2}\,.
\end{align}
In the quasiparticle approximation, i.e.~in the limit of va\-nishing 
self-energy, the diagonal spectral function \eqref{GFrhodiagSol} reverts 
to \eqref{GrhoQP}. From \eqref{GRdiagSol} we also obtain
\begin{align}
   \hat {\cal G}^{ii}_{h}(X,p)= -\frac{\Omega^{ii}(X,p)}
   {[\Omega^{ii}(X,p)]^2+{\textstyle\frac14}[\Pi^{ii}_\rho(X,p)]^2}\,.
\end{align}
In the quasiparticle approximation it reduces to \eqref{GhReal}. 
Combining \eqref{GrhoSol} and \eqref{GFsolution}  with 
\eqref{GRdiagSol} and \eqref{GFrhodiagSol1} we can express the 
full propagators in terms of the diagonal ones:

\begin{align}
   \label{GFrhoSolution}
   \hat G_{F(\rho)}&(X,p) =\bigl[\hat I +\hat {\cal G}_R (X,p) \hat \varPi^{'}_R
   (X,p)\bigr]^{-1}\nonumber\\
   &\times \bigl[\hat {\cal G}_{F(\rho)}(X,p)-\hat {\cal G}_R(X,p) \hat 
   \varPi^{'}_{F(\rho)}(X,p) \hat {\cal G}_A(X,p)\bigr]\nonumber\\
   &\times\bigl[\hat I + \hat \varPi^{'}_A(X,p) \hat {\cal G}_A
(X,p)\bigr]^{-1}\,,
\end{align}
where $\hat \varPi^{'}$ is the off-diagonal part of $\hat \Pi$. From 
\eqref{GgtrlessDef} and the definitions of the Wigner-transforms of the 
statistical propagator and spectral function it follows that the  
corresponding Wightman propagators are given by  
\begin{align}
   \hat G_\gtrless(X,p)=\hat G_F(X,p)\pm{\textstyle\frac{1}2}\, \hat
G_\rho(X,p)\,. 
\end{align}
Inverting the resulting (\textit{two-by-two}) matrices and taking into account
that the products of 
$\varPi^{'}_{R(A)}$ and ${\cal G}_{R(A)}$ are off-diagonal we arrive 
at \eqref{GgtrlessSolWT}.

\subsection{Boltzmann equation}

To obtain Boltzmann equations we neglect the (conventional) Poisson 
brackets on the right-hand side of \eqref{GFrhoQKeq}. In the 
Boltzmann approximation the field's mass disappears in the difference 
of the \textit{diagonal} components. However, the difference of 
the masses squared, $\Delta M^2_{ij}=M^2_i-M^2_j$,  appears in the 
equations for the off-diagonal terms. Moving these terms to the 
right-hand side we find
\begin{align}
   \label{GFBlzmn}
   p^\alpha{\cal D}_\alpha \hat{G}_{F(\rho)}
   = & {\textstyle\frac{i}2}([\hat \Omega_R\hat G_{F(\rho)}-\hat G_{F(\rho)}
\hat \Omega_A]\nonumber\\
     & -[\hat{\Pi}_{F(\rho)}\hat{G}_{A}-\hat{G}_{R}\hat{\Pi}_{F(\rho)}])\,,
\end{align}
where all the functions are evaluated at the same point $(X,p)$ 
of the phase-space. Substituting the equilibrium solutions 
\eqref{GrhoSol} and \eqref{GFsolution} into \eqref{GFBlzmn} we see 
that its right-hand side vanishes indeed.

If all the matrices in \eqref{GFBlzmn} were diagonal, it would 
revert to two independent systems of Boltzmann equations for 
the statistical propagator and spectral function:
\begin{subequations}
   \label{GFrhodiagBlzmn}
   \begin{align}
      \label{GFdiagBlzmn}
      p^\alpha{\cal D}_\alpha {\cal G}^{ii}_{F}&=
      {\textstyle\frac{1}2}[\,{\cal G}^{ii}_{F}\Pi^{ii}_{\rho}
      - \Pi^{ii}_{F}{\cal G}^{ii}_{\rho}\,]\,,\\
      \label{GrhodiagBlzmn}
      p^\alpha{\cal D}_\alpha {\cal G}^{ii}_{\rho}&=0\,.
   \end{align}
\end{subequations}
Equation \eqref{GrhodiagBlzmn} is consistent with the fact, that 
for a single scalar field the solution for the spectral function 
\eqref{GFrhodiagSol} is valid up to the first order in the gradients. 
This allows us to employ the Kadanoff--Baym ansatz for the 
statistical propagator, ${\cal G}_F^{ii}=(f_{\psi_i}+\frac12)
{\cal G}_\rho^{ii}$, and rewrite the original system of the 
Boltzmann equations \eqref{GFrhodiagBlzmn} as a Boltzmann equation 
for the distribution function $f_{\psi_i}$.
In the case of two mixing scalar fields the situation is quite 
different. The solution \eqref{GrhoSol} is valid only in equilibrium. 
As follows from \eqref{GFrhoconstraint},
out of equilibrium the off-diagonal components of the spectral 
function receive corrections linear in the gradients. Strictly speaking,
this means that we can not introduce a generalized Kadanoff--Baym
ansatz and reduce \eqref{GFBlzmn} to a system of two equations for
the one-particle distribution functions. In general, to analyze the 
generation of the asymmetry, one has to solve the system of the 
Boltzmann equations \eqref{GFBlzmn} for the full spectral function 
and statistical propagator. Only when the off-diagonal components 
are subdominant (which is the case in the hierarchical and resonant 
regime, but not in the maximal resonant limit) one can approximate 
\eqref{GFBlzmn} by \eqref{GFrhodiagBlzmn}.
The Boltzmann equations for the diagonal propagators 
\eqref{GFrhodiagBlzmn} do not contain the self-energy \CP-violating 
parameter. Given that the contributions proportional to the 
vertex \CP-violating parameter also cancel out in the Boltzmann 
equations for the real scalars \cite{Garny:2009rv}, this is an 
expected property.

% =============================================================================
\section{\label{ContourIntegration}Integration over the contour}
% =============================================================================

In this appendix we calculate two integrals over the
closed-time-path contour, shown in \fig\ref{contour}, which are required for
the derivation of the self-energy contribution to the \CP-violating parameter
in the Kadanoff--Baym formalism. Let us first consider the integral
\begin{align}
   I(x,y)={\textstyle \int\limits_{\cal C}} \dg u\, A(x,u) B(u,y)\,.
\end{align}
Assuming that the two-point functions possess a decomposition into
statistical and spectral components similar to \eqn\eqref{GfGr}, we find
\begin{align}
   \label{SingleIntFrho}
   i I_{F(\rho)}(x,y)=&{\textstyle \int\limits_{0}^{x^0}}
   \dg u \,A_\rho(x,u)B_{F(\rho)}(u,y)\nonumber\\
   &\hspace{4mm}-{\textstyle \int\limits_{0}^{y^0}} \dg u\, A_{F(\rho)}(x,u)
   B_\rho(u,y)\,.
\end{align}
These formulas are required to obtain the right-hand side of 
\eqref{KBeqs_real}. Replacing $A_\rho$ and $B_\rho$ by  
$A_R$ and $B_A$ we can rewrite the right-hand side of 
\eqref{SingleIntFrho} as an integral over the whole $u$
axis. From \eqref{SingleIntFrho} it also follows, that 
\begin{align}
   \label{SingleIntR}
   iI_{R}(x,y)=&\textint \dg u\, \theta(u^0) \,A_{R}(x,u)B_{R}(u,y)\,.
\end{align}
Next we consider the integral
\begin{align}
   J(x,y)={\textstyle \iint\limits_{\cal CC}} \dg u\, \dg v\, 
   A(x,u) B(u,v)C(v,y)\,.
\end{align}
Since $J(x,y)$ can be represented as an integral of a product of 
$I(x,v)$ and $C(v,y)$ over the contour, its spectral and statistical
components are given by \eqref{SingleIntFrho} with $A$ replaced 
by $I$. Furthermore, using \eqref{SingleIntR} we find
\begin{align}
   \label{DoubleIntFrho}
   J_{F(\rho)}(x,y)=-\,{\textstyle \iint}& \dg u \,  
   \dg v \,\theta(u^0)\,\theta(v^0) \nonumber\\
   \times [\,&A_R(x,u)B_R(u,v)C_{F(\rho)}(v,y)\nonumber\\
   +&A_R(x,u)B_{F(\rho)}(u,v)C_A(v,y)\nonumber\\
   +&A_{F(\rho)}(x,u)B_A(u,v)C_A(v,y)]\,.
\end{align}
Building the linear combinations of the spectral and 
statistical components we can easily derive the 
$\gtrless$ components from \eqref{DoubleIntFrho}.

We will now calculate the Wigner-transform of \eqref{DoubleIntFrho} 
in the Boltzmann approximation. That is, in each of the functions in
\eqref{DoubleIntFrho} we  neglect the deviation of the corresponding 
center coordinate from $X\equiv X_{xy}$. For instance:
\begin{align}
   A(x,u)\rightarrow A(X_{xu},s_{xu}) \rightarrow  A(X_{xy},s_{xu})\,.
\end{align}
In this approximation the integration over $u$ and $v$ induces a 
simple relation between the momenta $q_1=q_2=q_3\equiv q$. Integration 
over the relative coordinate $s$  induces an additional constraint 
$q=p$. Therefore we obtain:
\begin{align}
   \label{DoubleIntFrhoWT}
   J_{F(\rho)}(X,p)=- [\,&A_R(X,p)B_R(X,p)C_{F(\rho)}(X,p)\nonumber\\
   +&A_R(X,p)B_{F(\rho)}(X,p)C_A(X,p)\nonumber\\
   +&A_{F(\rho)}(X,p)B_A(X,p)C_A(X,p)]\,.
\end{align}
This completes the calculation of the Winger transform.

% =============================================================================
\section{\label{Ren}Renormalization}
% =============================================================================

In this appendix, we derive the renormalization prescription~\eqref{PiRen}
employed in Sec. \ref{CPviol} for the derivation of the \CP-violating
parameter in the resonant case.
For the level of approximation considered
there, it is sufficient to include perturbative one-loop mass and
field counterterms (see Appendix~\ref{cpclassic})%
\footnote{%
   The renormalization of the full Kadanoff--Baym equations~\eqref{KBeqs_real}
   has been discussed in refs.~\cite{Borsanyi:2008ar,Garny:2009ni},
   see also~\cite{vanHees:2001ik,vanHees:2002bv,Blaizot:2003an,Berges:2004hn,
   Berges:2005hc}.
}%
for the real scalar field $\psi_i$,
\begin{align}\label{lagrangianRenorm}
   \delta{\cal L} & = \frac12 \delta Z_{ij} \partial^\mu \psi_i \partial_\mu
\psi_j  -\frac12 \delta M^2_{ij} \psi_i\psi_j \,.
\end{align}
Including the counterterms in the Lagrangian~\eqref{lagrangian}
results in a modification of the Schwinger-Dyson
equation~\eqref{SchwingerDyson},
\begin{align}\label{SchwingerDysonRen}
   \hat G^{-1}(x,y)= \mathscr{\hat G}^{-1}(x,y) - \hat\Pi^{ren}(x,y)\,,
\end{align}
(using matrix notation; see Appendix \ref{SD}), where
\[
   \hat\Pi^{ren}(x,y) = \hat\Pi(x,y) - i (\delta\hat Z \Box_x + \delta\hat M^2 )
 \delta^g(x,y) \;.
\]
Proceeding analogously as in Sec. \ref{CPviol}, we decompose the renormalized
self-energy into diagonal and off-diagonal parts,
\[
   \hat \Pi^{ren}(x,y) = \hat \varPi^{ren}(x,y) + (\hat \varPi^{ren})^{'}(x,y)
\,,
\]
and define a renormalized ``diagonal'' propagator by
\begin{align}\label{DiagonalEqRen}
   [\hat {\cal G}^{ren}]^{-1}(x,y) = \hat{\mathscr{G}}^{-1}(x,y) - \hat
\varPi^{ren}(x,y)\,.
\end{align}
In analogy to the steps leading to Eq.\,\eqref{GgtrlessFormalSol}, we find for
the full Wightman functions
\begin{widetext}
\begin{eqnarray}\label{GgtrlessFormalSolRen}
   \hat G_\gtrless(x,y) & = & \hat {\cal G}_\gtrless^{ren}(x,y) -
{\textstyle\iint} \mathscr{D}^4u \mathscr{D}^4v\, \theta(u^0) \, \theta(v^0)
                              \bigl[\hat G_R(x,u) \hat \varPi^{'}_\gtrless(u,v)
\hat {\cal G}^{ren}_A(v,y)
                              + \hat G_\gtrless(x,u)\hat \varPi^{'}_A(u,v)\hat
{\cal G}^{ren}_A(v,y) \nonumber\\
                        &   & \qquad\qquad\qquad\qquad\qquad\qquad\qquad\qquad
                              + \hat G_R(x,u)\hat \varPi^{'}_R(u,v)\hat {\cal
G}^{ren}_\gtrless(v,y)\bigr] \\
                        &   & {} - {\textstyle\int} \mathscr{D}^4u \;
\theta(u^0)
                              \bigl[ \hat G_\gtrless(x,u) (\delta\hat Z' \Box_u
+ (\delta\hat M^2)') \hat {\cal G}^{ren}_A(u,y)
                              + \hat G_R(x,u) (\delta\hat Z' \Box_u +
(\delta\hat M^2)')  \hat {\cal G}^{ren}_\gtrless(u,y)\bigr] \; , \nonumber
\end{eqnarray}
\end{widetext}
where $\delta\hat Z'$ and  $( \delta\hat M^2)'$ denote the off-diagonal parts of
$\delta\hat Z$ and  $\delta\hat M^2$, respectively.
Using this, we find that the renormalized version of Eq.\,\eqref{GgtrlessSolWT}
can be obtained by replacing $\hat {\cal G} \rightarrow \hat {\cal G}^{ren}$,
as well as $\hat\varPi'_{R(A)}(X,p) \rightarrow \hat\varPi'_{R(A)}(X,p) -
\delta\hat Z'p^2 + ( \delta\hat M^2)'$.
The latter prescription coincides with Eq.\,\eqref{PiRen} for $i\not= j$.

The diagonal part of the self-energy enters in the renormalized
diagonal propagator $\hat {\cal G}^{ren}$; see Eq.\,\eqref{DiagonalEqRen}.
Since this equation can be split into independent equations for each entry on
the diagonal, it is analogous to the case of a single real scalar field; see
e.g.~\cite{Hohenegger:2008zk}. The solution of the corresponding
Wigner-transformed kinetic equations for the retarded and advanced propagator
reads (up to second-order gradients; see also Eq.\,\eqref{GRdiagSol}),
\begin{equation}\label{GRetRenorm}
   \mathcal{G}_{R(A)}^{ren,ii}(X,p) = - [ p^2 - M_i^2 - \Pi_{R(A)}^{ren,
ii}(X,p) ]^{-1}\;,
\end{equation}
where the renormalized self-energies are determined in accordance with
Eq.\,\eqref{PiRen} for $i = j$. Since $\mathcal{G}_{\rho}^{ren}(X,p) =
2\Im\mathcal{G}_{R(A)}^{ren}(X,p)$, the spectral function also contains
renormalized self-energies. Furthermore, using the Kadanoff--Baym ansatz
${\cal G}_F^{ren,ii}=(f_{\psi_i}+\frac12){\cal G}_\rho^{ren,ii}$ implies that
also the statistical propagator, and therefore also the Wightman functions, are
finite.

Thus, altogether, we find that the prescription~\eqref{PiRen} renormalizes the
diagonal as well as off-diagonal components of the relevant propagators.

% =============================================================================
\section{\label{FullKB}Analysis in the full KB formalism}
% =============================================================================

In this appendix, we derive a time-evolution equation for the
``baryon'' asymmetry within nonequilibrium field theory. It is based on
the approximate $U(1)_B$-symmetry $b \rightarrow e^{i\alpha}b$ of the
toy-model Lagrangian~(\ref{lagrangian}),  and does not require any
further approximations beyond the 2PI truncation. As we will see,
in the Boltzmann limit, the resulting equation for the asymmetry coincides
with those obtained from the quantum-corrected Boltzmann equations discussed
above.

The Lagrangian~(\ref{lagrangian}) can be split into $B$-conserving and
$B$-violating parts,
\[
   {\cal L} = {\cal L}_B + {\cal L}_{\Delta B} \;,
\]
where
\begin{align}
   {\cal L}_B          & = {\cal L}_{kin} -\frac12 M^2_i \psi_i\psi_i -
m^2\bar{b}b -\frac{\lambda}{2!2!}(\bar{b}b)^2 \;, \\
   {\cal L}_{\Delta B} & = {} - \frac{g_i}{2!}\psi_i bb - \frac{g^*_i}{2!}\psi_i
\bar{b}\bar{b} \;.
\end{align}
The corresponding Noether current is given by
\[
   J_\mu(x) = -2i\left( b(x) \partial_\mu \bar b(x) - \partial_\mu b(x) \bar
b(x) \right) \;.
\]
Its expectation value can be written as (note that we are working in the
Heisenberg picture)
\[
   j_\mu(x) = \langle J_\mu(x) \rangle = 2i \left( \partial_{x^\mu} -
\partial_{y^\mu} \right) D_F(x,y) |_{x=y} \;,
\]
where $D_F(x,y) = \frac12 \langle b(x)\bar b(y) + \bar b(y)  b(x) \rangle$ 
is the statistical propagator of the complex $b$-field.

Because of the presence of ${\cal L}_{\Delta B}$, the Noether current is in general
not conserved. Its divergence reads
\begin{align}\label{DivJ}
   \partial^\mu j_\mu(x) & = {\textstyle \frac12} \left. \left( \partial_{x_\mu}
+ \partial_{y_\mu} \right) \left[ 2i \left( \partial_{x^\mu} - \partial_{y^\mu}
\right) D_F(x,y) \right] \right|_{x=y} \nonumber \\
                         & = i \left. \left[ \Box_x - \Box_y \right] D_F(x,y)
\right|_{x=y} \;.
\end{align}
The expression in the last line can be evaluated by using the
Kadanoff--Baym equations for the complex scalar field.
These can be derived analogously to the system \eqref{KBeqs_real} 
of Kadanoff--Baym equations for the real scalar fields. They read
\begin{subequations}
   \label{Dequations}
   \begin{align}
      \label{DFequation}
      [\square_x+m^2(x)]D_F(x,y)&=
      {\textint\limits^{y^0}_0} \dg z\,\Sigma_F(x,z)D_\rho(z,y)\nonumber\\
      &-{\textstyle\int\limits^{x^0}_0}\dg z\,\Sigma_\rho(x,z)D_F(z,y)\\
      \label{Drhoequation}
      [\square_x+m^2(x)]D_\rho(x,y)&=
      {\textint\limits_{x^0}^{y^0}}\dg z\,\Sigma_\rho(x,z)D_\rho(z,y)\,.
      \hskip -1mm
   \end{align}
\end{subequations}
We use the notations of Ref.~\cite{Garny:2009rv} here.
In particular, the statistical propagator $D_F$ and the spectral function
$D_\rho$ of the complex field are defined analogously to Eq.\,(\ref{GFGrhoij}).
The upper equations possess an equivalent representation in terms of the
antiparticle propagators
\begin{align}\label{DbarDef}
   \bar D_F(x,y)    & \equiv D_F(y,x) \; , \nonumber\\
   \bar D_\rho(x,y) & \equiv - D_\rho(y,x) \; ,
\end{align}
and antiparticle self-energies $\bar \Sigma_{F,\rho}$ (defined analogously).
The Kadanoff--Baym equations for the antiparticles then take the same form as
Eqs.~(\ref{Dequations}), except for the replacement $D \rightarrow \bar D$ and
$\Sigma \rightarrow \bar \Sigma$.

The Kadanoff--Baym equations for $D_F$ and $\bar D_F$ can now be
used to evaluate the right-hand side of Eq.\,(\ref{DivJ}),
\begin{eqnarray}\label{DivJ_KB_Derivation}
   \lefteqn{ \partial^\mu j_\mu(x) = i \left. \left[ \Box_x + m^2(x)
\right]\left[  D_F(x,y) - \bar D_F(x,y) \right] \right|_{x=y} } \nonumber \\
        & = & i \textint\limits^{x^0}_0 \dg z\, \bigl\{ \Sigma_F(x,z)D_\rho(z,x)
- \Sigma_\rho(x,z)D_F(z,x) \nonumber\\
        &   & {} - \bar\Sigma_F(x,z)\bar D_\rho(z,x) + \bar \Sigma_\rho(x,z)\bar
D_F(z,x) \bigr\} \;.
\end{eqnarray}
By using the retarded and advanced propagators
\begin{eqnarray}
   D_R(x,y) & = & \Theta(x^0-y^0)D_\rho(x,y) \;, \nonumber\\
   D_A(x,y) & = & -\Theta(y^0-x^0)D_\rho(x,y) \;,
\end{eqnarray}
as well as analogous definitions for the antiparticle propagators, we finally
find the following equation of motion for the toy-baryon current:
\begin{eqnarray}\label{DivJ_KB}
   \partial^\mu j_\mu(x) & = & -i \textint \dg z\, \Theta(z^0) \times \\
   &   & {} \times \bigl\{ \Sigma_F(x,z)D_A(z,x) + \Sigma_R(x,z)D_F(z,x)
\nonumber \\
   &   & {} - \bar\Sigma_F(x,z)\bar D_A(z,x) - \bar \Sigma_R(x,z)\bar D_F(z,x)
\bigr\} \nonumber\;.
\end{eqnarray}
So far, the derivation is quite general and the upper equations are formally
fulfilled even in the exact theory.
To obtain a closed system we have to express the self-energies in terms of 
the two-point functions. In the Kadanoff--Baym formalism the self-energies
are obtained by functional differentiation of the 2PI effective action.
Approximations can be obtained by truncating the 2PI functional, for example
at a certain loop order. This leads to so-called $\Phi$-derivable
approximations,
which are conserving, i.e.~they automatically respect the conservation laws of
the full theory~\cite{Ivanov:1998nv,Knoll:2001jx}. This means that, in the
limit of vanishing $B$-violation, the baryon current obtained in the
Kadanoff--Baym formalism is indeed conserved at any loop order of the 2PI
functional.

The structure  of the 2PI effective action in two-loop approximation can 
be read off the diagram in Fig.\,\ref{diagrams} (a):
\begin{align}
   i\Gamma_2 = & -{\textstyle\frac14}\, g_m g_n^*\intg \dg x \dg y\, 
                 G^{mn}(x,y) D^2(x,y)\nonumber\\
               & -{\textstyle\frac14}\, g^*_m g_n \intg \dg x \dg y\,
                 G^{mn}(x,y) D^2(y,x)\,.
\end{align}
Differentiating the effective action with respect to the two-point 
function of the complex scalar field we obtain
\begin{align}
   \label{Sigma}
   \Sigma(x,y) & = -g^*_i g_jG_{ij}(x,y)D(y,x)\,.
\end{align}
The statistical and spectral components of the self-energy 
are given by
\begin{subequations}
   \label{SigmaTwoLoop}
   \begin{align}
      \Sigma_F(x,y)=-g^*_i g_j [\,G_F^{ij}&(x,y)D_F(y,x)\nonumber\\
      &+{\textstyle\frac14}\,G_\rho^{ij}(x,y)D_\rho(y,x)]\,,\\
      \Sigma_\rho(x,y)=-g^*_i g_j [\,G_F^{ij}&(x,y)D_\rho(y,x)\nonumber\\
      &-\,G_\rho^{ij}(x,y)D_F(y,x)]\,.
   \end{align}
\end{subequations} 
The upper expressions, together with the Kadanoff--Baym 
equations~(\ref{Dequations}), as well as the corresponding equations 
for the toy-Majorana fields~(\ref{KBeqs_real}) 
and~(\ref{FullPi}), form a closed system of equations. Inserting 
their solutions into Eq.\,(\ref{DivJ_KB}) yields the time-evolution 
of the toy-baryon asymmetry
\begin{equation}\label{nB_Def}
   n_B(t) \equiv \frac{1}{V} \textint d^3x \; j_0(t,\vec{x}) \;.
\end{equation}
Equation (\ref{DivJ_KB}) takes memory and quantum off-shell effects into account
in a very general way, and is valid even very far from thermal equilibrium.
Furthermore, due to the self-consistency it implicitly resums a large number of
multiparticle scattering processes. However, this feature also makes the
(numerical) solution of the full equations difficult.
In a typical leptogenesis scenario one is interested in a situation where the
deviations from equilibrium are comparably mild. In this case, it is possible
to obtain simplified evolution equations for the asymmetry from Eq.\,(\ref{DivJ_KB}). 
In the following, we will show that the resulting equations are equivalent to 
those obtained from the kinetic evolution equations of the particle distribution
function, provided both are based on the identical truncation
of the 2PI effective action.

\subsection{Consistency with the kinetic approach}

Usually, within the kinetic approach, the baryon asymmetry is
calculated by evaluating the difference of particle and antiparticle
distribution functions,
\begin{equation}
   n_b(t) - n_{\bar b}(t) = {\textstyle \int \frac{d^3k}{(2\pi)^3}} \left[
f_b(t,\vec{k}) - f_{\bar b}(t,\vec{k}) \right] \;,
\end{equation}
where $f_b(t,\vec{k})$ and $f_{\bar b}(t,\vec{k})$ are obtained from
Boltzmann-like kinetic equations. Equations (\ref{nB_Def},\ref{DivJ_KB}) for
the toy-baryon asymmetry discussed above provide a complementary possibility to
calculate the asymmetry. In this context, the question arises, under which
conditions both approaches are equivalent. In the following, we will show that
equivalence holds provided (i) one uses the \emph{top-down} or
\emph{quantum-corrected} kinetic equations, which are obtained from the
gradient expansion of the full Kadanoff--Baym equations, (ii) the evolution
equation~(\ref{DivJ_KB}) of the baryon current is expanded to the same order in
the gradient expansion, and (iii) both sets of equations are based on the
same truncation of the 2PI effective action. This can be seen as follows: By
performing a Wigner transformation and a first-order gradient expansion of
Eq.\,(\ref{DivJ_KB}), and using relations~(\ref{DbarDef}), we obtain on the
one hand
\begin{eqnarray}\label{DivJ_GradExp}
   \partial^\mu j_\mu(X) & = & - {\textstyle \int\frac{d^4p}{(2\pi)^4}} \bigl[
\Sigma_F(X,p)D_\rho(X,p) \nonumber\\
                         &   & {} - \Sigma_\rho(X,p)D_F(X,p) -
\{\Sigma_F,D_h\}_{PB}  \nonumber\\
                         &   & {} - \{\Sigma_h,D_F\}_{PB} \bigr]\;,
\end{eqnarray}
where the curly brackets denote the usual Poisson brackets, and we have used
the decomposition $D_{R(A)}(X,p) = D_h(X,p) \pm \frac{i}{2}D_\rho(X,p)$.

On the other hand, the quantum kinetic equation for $D_F$ obtained from a
first-order gradient expansion of the Kadanoff--Baym
equation~(\ref{DFequation}) reads (see e.g.~Ref.~\cite{Garny:2009rv})
\begin{eqnarray}
   \lefteqn{ \{\omega(X,p),D_F(X,p)\}_{PB} = \Sigma_\rho(X,p)D_F(X,p) }
\nonumber\\
   && - \Sigma_F(X,p)D_\rho(X,p) + \{\Sigma_F,D_h\}_{PB} \;,
\end{eqnarray}
where $\omega(X,p)=p^2-m^2-\Sigma_h(X,p)$. Inserting this equation into
Eq.\,(\ref{DivJ_GradExp}) yields
\begin{equation}
   \partial^\mu j_\mu(X)  =  {\textstyle \int \frac{d^4p}{(2\pi)^4}} \;
2p^\alpha {\cal D}_\alpha D_F(X,p) \;.
\end{equation}
Using $\bar D_F(X,p) = D_F(X,-p)$, we can write this as
\begin{equation}
   \partial^\mu j_\mu(X)  =  {\textstyle \int \!\! \frac{d^3p}{(2\pi)^3}
\int\limits_0^\infty \! \frac{dp^0}{2\pi}} \, 2p^\alpha {\cal D}_\alpha \left[
D_F(X,p) - \bar D_F(X,p)\right].
\end{equation}
Inserting the Kadanoff--Baym ansatz
$D_F(X,p)=\bigl[f_b(X,p)+\frac12\bigr]D_\rho(X,p)$ yields
\begin{equation}
   \partial^\mu j_\mu(X)  = {\textstyle  \int \!\! \frac{d^3p}{(2\pi)^3}
\int\limits_0^\infty \!\! \frac{dp^0}{2\pi}} \, 2p^\alpha {\cal D}_\alpha \left[
f_b - f_{\bar b} \right] D_\rho(X,p),
\end{equation}
where we have assumed a symmetric spectrum, $D_\rho=\bar D_\rho$. For a
spatially homogeneous system, the frequency integration yields the particle
distribution function of the momentum mode $\vec{p}$ at time $t=X^0$,
\[
   f_b(X^0,\vec{p}) \equiv {\textstyle \int\limits_0^\infty \! \frac{dp^0}{2\pi}
}\, 2p^0 f_b(X,p) D_\rho(X,p) \;.
\]
Thus, using Eq.\,(\ref{nB_Def}), we find
\begin{eqnarray}
   \frac{d}{dt} n_B(t) & = & \frac{1}{V} \textint d^3x \, \partial_t
j_0(t,\vec{x}) \nonumber\\
                       & = & {\textstyle \int \!\! \frac{d^3p}{(2\pi)^3}}\,
\partial_t \left( f_b(t,\vec{p}) - f_{\bar b}(t,\vec{p}) \right) \nonumber\\
                       & = & \frac{d}{dt} \left( n_b(t) - n_{\bar b}(t)
\right)\;.
\end{eqnarray}
Thus, we find that the baryon asymmetry $n_B(t)$ inferred from the equation of
motion of the baryon current coincides with the asymmetry
$n_b(t) - n_{\bar b}(t)$ obtained from the corresponding kinetic approach,
under the conditions discussed above (this is in accordance with the general
analysis of Ref.~\cite{Knoll:2001jx}).

\subsection{Self-energy contribution}

Here, we briefly want to discuss the time-evolution of the baryon asymmetry
within the 2PI two-loop truncation based on the full evolution
equation~(\ref{DivJ_KB}).
As discussed in Sec.\,\ref{kinEq}, the 2PI two-loop approximation captures
the self-energy contribution to the CP-violating decays and inverse decays,
whereas the vertex contribution requires to take also three-loop diagrams
in the 2PI functional into account~\cite{Garny:2009rv}. Thus, the 2PI two-loop
case provides a possibility to investigate whether an asymmetry can be induced
by the self-energy-type contributions. The latter question has, after some
controversial discussions, been answered e.g.~in Ref.~\cite{Roulet:1997xa}
within the Boltzmann framework. In the following, we address the same
question, from a slightly different point of view, within nonequilibrium
quantum field theory.

Assuming spatial homogeneity, we can write the total baryon asymmetry as
\begin{equation}
   n_B(t) = \frac{1}{V} \textint d^3x \, j_0(t,\vec{x})  \equiv 
        {\textstyle \int \frac{d^3k}{(2\pi)^3}}\, n_B(t,\vec{k}) \;.
\end{equation}
From Eq.\,(\ref{DivJ_KB}), we obtain the following evolution equation for the
baryon asymmetry in momentum mode $\vec{k}$:
\begin{eqnarray}
   \lefteqn{ \frac{d}{dt} n_B(t,\vec{k}) \ = } \\
   &   & i \textint\limits^{t}_0 dt' \bigl\{
\Sigma_F(t,t',\vec{k})D_\rho(t',t,\vec{k}) -
\Sigma_\rho(t,t',\vec{k})D_F(t',t,\vec{k}) \nonumber\\
        &   & {} - \bar\Sigma_F(t,t',\vec{k})\bar D_\rho(t',t,\vec{k}) + \bar
\Sigma_\rho(t,t',\vec{k})\bar D_F(t',t,\vec{k}) \bigr\} \;, \nonumber
\end{eqnarray}
where $D(x^0,y^0,\vec{k})\equiv \int d^3x e^{i\vec{k}(\vec{x}-\vec{y})} D(x,y)$.

By inserting the 2PI two-loop expressions~(\ref{SigmaTwoLoop}) into the upper
equation, and using Eq.\,(\ref{GFGrhoProp}), we obtain
\begin{eqnarray}\label{nB_TwoLoop}
   \lefteqn{ \frac{d}{dt} n_B(t,\vec{k}) \ = } \nonumber \\
   &  & -2i \textint\limits^{t}_0 dt' \bigl\{ \Re(g_ig_j^*) \bigl[ 2G_F^{ij}
\bigl( D_FD_\rho - \bar D_F\bar D_\rho \big)  \nonumber \\
   &  & {} + G_\rho^{ij} \big( D_F^2 - \bar D_F^2 - {\textstyle \frac{1}{4}} (
D_\rho^2 - \bar D_\rho^2 ) \big) \bigr] \nonumber\\
        &  & {} + i \Im(g_ig_j^*) \bigl[ 2G_F^{ij} \big( D_FD_\rho + \bar
D_F\bar D_\rho \bigr)  \nonumber \\
   &  & {} + G_\rho^{ij} \big( D_F^2 + \bar D_F^2 - {\textstyle \frac{1}{4}} (
D_\rho^2 + \bar D_\rho^2 ) \bigr) \bigr] \bigr\} \;,
\end{eqnarray}
where all propagators are evaluated at $(t',t,\vec{k})$. In the limit of
vanishing CP-violation, i.e.~for $\Im(g_ig_j^*)=0$, the Eq.\,\eqref{nB_TwoLoop}
possesses a baryon-symmetric solution for which $D_F=\bar D_F$,
$D_\rho=\bar D_\rho$ (see Appendix~\ref{SymmConf}), and $n_B=0$ at all times.
For an initial state which is asymmetric, the terms proportional to
$\Re(g_ig_j^*)$ tend to wash out the asymmetry and would, in flat space-time,
asymptotically drive $n_B \rightarrow 0$.

Here, we are interested in the opposite situation. This means, we want to
answer the question whether $n_B(t)$ can deviate from zero at some instant of
time \emph{although} it vanishes initially, $n_B(t=0)=0$. Any change of $n_B$
with time requires, of course, nonzero B-violation (i.e.~$g_i\not= 0$), since
$n_B$ would correspond to a conserved charge otherwise. Additionally, from
Eq.\,(\ref{nB_TwoLoop}), we see that the generation of an asymmetry also
requires $\Im(g_ig_j^*) \not= 0$, i.e.~nonvanishing \CP-violation. Finally, a
deviation from thermal equilibrium is required, since otherwise
time-translation invariance would require $n_B(t)$ to be equal to its initial
value at all times. Thus, the quantum evolution equation~(\ref{nB_TwoLoop})
obtained from the 2PI two-loop approximation is in accordance with the
Sakharov conditions. In general, we therefore expect that $n_B(t)$ can deviate
from $n_B(0)=0$ for some $t>0$, if $\Im(g_ig_j^*) \not= 0$ and if the system
deviates from equilibrium%
\footnote{%
   There are two potential sources for deviations from thermal equilibrium,
   namely the space-time expansion as well as the initial conditions. In order
   to check whether Eq.\,(\ref{nB_TwoLoop}) can indeed describe the generation
   of an asymmetry, we concentrate on the latter source for simplicity.
   We note that, in order to describe equilibrium initial conditions of the
   full quantum evolution equations, it would be necessary to include
   non-Gaussian correlations of the initial state (see e.g.~\cite{Garny:2009ni}).
   However, since the latter play no role in the kinetic approach (the
   initial time is formally sent to the infinite past when performing a Wigner
   transformation), we do not consider them here.
}.
In order to make this statement more quantitative, we consider a
Taylor expansion of Eq.\,(\ref{nB_TwoLoop}) around the initial time $t=0$. The
first nonvanishing contribution turns out to be of order four,
\begin{eqnarray}
   \lefteqn{ \frac{d^4}{dt^4} n_B(t,\vec{k})|_{t=0} \ = } \nonumber \\
   &  & {} -4\Im(g_1g_2^*)D_F\partial_t(G^{12}-G^{21})|_{t=t'=0} \\
   &  & {} -2n_B(0,\vec{k})(\Re(g_ig_j^*)G_F^{ij}+|g_i|^2D_F)|_{t=t'=0}
\nonumber\;.
\end{eqnarray}
Thus, for a symmetric initial state with $n_B(0,\vec{k}) = 0$, an asymmetry can
arise if $\Im(g_1g_2^*) \not= 0$, as expected, and if
$\partial_t(G^{12}-G^{21})|_{t=t'=0}\not=0$. The latter condition is, however,
not necessary. If $\partial_t(G^{12}-G^{21})|_{t=t'=0}=0$ and $n_B(0,\vec{k}) =
0$,
the leading term reads
\[
   \frac{d^5}{dt^5} n_B(t,\vec{k})|_{t=0} =
4\Im(g_1g_2^*)(M_2^2-M_1^2)D_FG_F^{12}|_{t=t'=0} \;.
\]
Consequently, the full quantum evolution equations support the observation that
an asymmetry can be generated from self-energy-type contributions to the
\CP-violating decays.

% =============================================================================
\section{\label{SymmConf}Baryonically symmetric configuration}
% =============================================================================

The Wightman propagators of the complex scalar field are defined by
\cite{Garny:2009rv} 
\begin{subequations}
   \label{gtrlesscomponents}
   \begin{align}
      D_>(x,y)&\equiv\langle b(x)\bar{b}(y)\rangle=
      \Tr[\mathscr{P}\,b(x)\bar{b}(y)]\,,
      \\
      D_<(x,y)&\equiv\langle \bar{b}(y) b(x)\rangle=
      \Tr[\mathscr{P}\,\bar{b}(y)b(x)]\,.
   \end{align}
\end{subequations}
Applying the operator of charge conjugation $C$ to \eqref{gtrlesscomponents} 
we obtain
\begin{subequations}
   \label{CCgtrless}
   \begin{align}
      C D_>(x,y) C^{-1}&=
      \Tr[\mathscr{P}^c\,\bar b(x) b(y)]=D^c_<(y,x)\,,
      \\
      C D_<(x,y) C^{-1}&=\Tr[\mathscr{P}^c\,b(y)\bar{b}(x)]=
                D^c_>(y,x) \,,
   \end{align}
\end{subequations}
where the superscript `c' denotes the charge-conjugated quantities. 
In a baryonically symmetric configuration $\mathscr{P}^c=\mathscr{P}$. 
Therefore in this case $D_\gtrless(x,y)=D_\lessgtr(y,x)$. In turn, for 
the statistical propagator and spectral function this implies
\begin{align}
   D_F(x,y)= D_F(y,x)\,,\quad D_\rho(x,y)=- D_\rho(y,x)\,.
\end{align}
Since the full propagator is related to $D_F$ and $D_\rho$ by
\begin{align}
   \label{Ddecompos}
   D(x,y)&=  D_F(x,y)-{\textstyle\frac{i}{2}}\,\sign_{\cal C}(x^0-y^0)
D_\rho(x,y)\,,
\end{align}
we conclude that in a baryonically symmetric configuration 
$\bar D(x,y)\equiv D(y,x)=D(x,y)$\,.

Differentiating the effective action with respect to the two-point 
function of the real scalar fields we obtain the two-loop contribution 
to the self-energy: 
\begin{align}
   \label{Pi}
   \Pi^{ij}(x,y)&=-{\textstyle\frac12}g_i g^*_j D^2(x,y)
   -{\textstyle\frac12}g^*_i g_j \bar D^2(x,y)\,.
\end{align}
In the symmetric configuration it takes the form
\begin{align}
   \label{PiSymm}
   \Pi^{ij}(x,y)&=-{\textstyle\frac12}(g_i g^*_j+ g^*_i g_j) D^2(x,y)\,,
\end{align}
i.e.~this matrix is symmetric. This is also true for the spectral, statistical,
retarded and advanced self-energies that can be derived from \eqref{PiSymm}.

% =============================================================================
\section{\label{OneLoopInt}One-loop integral}
% =============================================================================

Here we calculate the one-loop self-energy integrals given in Eq.\,(\ref{LRhoDef}),
\begin{equation}
   L_{h(\rho)}(X,p) = 16\pi {\textstyle \int} d\Pi_k
D_F(X,k)D_{h(\rho)}(X,p-k)\,,
\end{equation}
evaluated for an on-shell momentum $p=(E_i,\vec{p})$ where
$E_i=(\vec{p}^2+M_i^2)^{1/2}$. We have omitted the superscript `s' here since
the calculation applies also to baryonically asymmetric states. We first
consider $L_\rho$. Inserting the quasiparticle approximation for the complex
field, see Eq.\,(\ref{DrhoQP}), and the Kadanoff--Baym ansatz
$D_F=\left(\frac{1}{2} + f_b\right) D_\rho$, one finds
\begin{align}
   L_\rho(X,p) = & 16\pi \! \textint \! d\Pi_k \left[{\textstyle \frac{1}{2}} +
f_b(k\cdot u) \right] 2\pi{\rm sgn}(k_0) \delta(k^2-m^2) \nonumber\\
                 & {} \times 2\pi {\rm sgn}(p_0-k_0) \delta((p-k)^2-m^2) \,,
\end{align}
where we have used a manifestly Lorentz-covariant notation. In the rest-frame
of the medium $u=(1,0,0,0)$. Note that only the poles with $k_0>0$ and
$p_0-k_0>0$ contribute since $p_0>M_i>2m$. The integration over $k_0$ is
trivial due to the Dirac-$\delta$ in the first line. In order to perform the
remaining integrations, we express the integrand in the rest-frame of the
decaying ``heavy neutrino'', for which $p'=(M_i,\vec{0})$ and
$u'=(E_i,-\vec{p})/M_i$,
\begin{equation}
   L_\rho(X,p) = 16\pi \! {\textstyle \int \! \frac{d^3k}{(2\pi)^3}}
\left[{\textstyle \frac{1}{2}}+ f_b(k\cdot u') \right]
\frac{\pi}{2\omega_k^2}\delta(M_i-2\omega_{k}) \,,
\end{equation}
where $\omega_k=(\vec{k}^2+m^2)^\frac12$ and
$k \cdot u' = (\omega_kE_i + \vec{kp})/M_i$.
The Dirac $\delta$-function requires $\omega_{k}=M_i/2$.
Using furthermore $\vec{kp}=|\vec{k}||\vec{p}|\cos\Theta$, we find
\begin{equation}
   k \cdot u' = (E_i + r|\vec{p}|\cos\Theta)/2 \equiv E_p \,,
\end{equation}
where $r=(1-4m^2/M_i^2)^{1/2}$. After integrating over $|\vec{k}|$, one obtains
\begin{align}
L_\rho(X,p)= r {\textstyle \int\frac{d\Omega}{4\pi}}[1+2f_b(E_p)]\,.
\end{align}
If the medium is approximately symmetric, $f_b \simeq f_{\bar b}$, we can write
the result in the form presented in Eq.\,(\ref{LRho}).

Let us now turn to $L_h$. As above, we insert the Kadanoff--Baym ansatz for
$D_F$ and use the quasiparticle approximation for $D_\rho$, yielding
\begin{align}
\hspace{-1mm}
   L_{h}(X,p) & = 16\pi \! \textint \! d\Pi_k \frac{\pi}{\omega_k}
\delta(k^0-\omega_k) \bigl\{ \left[{\textstyle \frac12} + f_b(k\cdot u)
\right]\\
              &  \times D_{h}(X,p-k) + \left[ {\textstyle \frac12} + f_{\bar
b}(k\cdot u) \right] \bar D_{h}(X,p+k) \bigr\}.\nonumber
\end{align}
In the quasiparticle limit, one can express $D_h\equiv\Re D_R$ in the form
\begin{equation}
	\label{D_h}
   D_h(X,p) = \bar D_h(X,p) = -\frac{\mathcal{P}}{p^2-m^2}\;,
\end{equation}
where $\mathcal{P}$ denotes the principal value. As above, we evaluate the
integral in the rest-frame of the decaying particle. The result of the 
calculation reads
\begin{align}
   L_h(X,p) = & \int \!\! \frac{d^3k}{(2\pi)^3} \frac{8\pi}{\omega_kM_i} \Big\{
\left[ {\textstyle \frac{1}{2}} + f_b(k\cdot u') \right] \frac{ \mathcal{P} }{
M_i-2\omega_{k} } \nonumber \\
               {} +& \left[ {\textstyle \frac{1}{2}} + f_{\bar b}(k\cdot u')
\right] \frac{ \mathcal{P} }{ M_i+2\omega_{k} } \Big\} \nonumber\\
       \equiv &  L_h^{vac}(p^2) + L_h^{med}(X,p)\;.
\end{align}
The vacuum contribution, which corresponds to the limit
$f_b,f_{\bar b} \rightarrow 0$, depends only on $p^2$ due to Lorentz invariance
of the vacuum state. It is logarithmically divergent. However, after
renormalization, only the difference $L_h^{vac}(p^2)-L_h^{vac}(\mu^2)$ appears,
which is finite. For the on-shell momentum $p^2=M_i^2$ considered here, we
obtain
\begin{align}
   L_h^{vac}(M_i^2) = & -\frac{1}{\pi}\left[ r \ln\left( \frac{rM_i}{2m} \right)
- r_\mu \ln\left( \frac{r_\mu M_i}{2m} \right) \right] \nonumber \\
               \rightarrow & - \frac{1}{2\pi} \ln \frac{M_i^2}{\mu^2} \quad
\mbox{for} \ m \rightarrow 0\;,
\end{align}
where  $r_\mu=(1-4m^2/\mu^2)^{1/2}$. For an approximately symmetric state, the
medium contribution can be written as
\begin{align}
   L_h^{med}(X,p) = & - 8\pi \int \!\! \frac{d^3k}{(2\pi)^3}
\frac{f_b(E)+f_{\bar b}(E)}{\omega_k} \frac{\mathcal{P}}{M_i^2-4\omega_k^2} \;,
\end{align}
where $E \equiv k\cdot u' \rightarrow
\frac{|\vec{k}|}{M_i}[ E_i + |\vec{p}|\cos\Theta ]$ in the limit
$m \rightarrow 0$. By introducing spherical coordinates and substituting
$|\vec{k}|$ for $E$, we obtain
\begin{align}
   L_h^{med}(X,p) = & - \frac{4}{\pi}\int_0^\infty \!\! dE E (f_b(E)+f_{\bar
b}(E)) \times \nonumber \\
                         & \times \int \frac{d\Omega}{4\pi}
\frac{\mathcal{P}}{(E_i+|\vec{p}|\cos\Theta)^2-4E^2} \;.
\end{align}
After performing the angular integration (and replacing
$M_i \rightarrow M_i^{med}$), we obtain the result stated in Eq.\,\eqref{Lh}.

\end{appendix}

% =============================================================================
%\bibliographystyle{apsrev}
%\bibliography{../references}
% =============================================================================

\end{document}